\documentclass[nofootinbib,superscriptaddress]{revtex4}
\usepackage{amsmath}
\usepackage{mathtools}
\usepackage{amssymb}
\usepackage{cancel}
\usepackage{amsfonts}
\usepackage[T1]{fontenc}
\usepackage{bm}
\usepackage{color}
\usepackage{graphicx}
\DeclareSymbolFont{operators}{OT1}{cmr}{m}{n}
\DeclareSymbolFont{letters}{OML}{cmm}{m}{it}
\DeclareSymbolFont{symbols}{OMS}{cmsy}{m}{n}
\DeclareSymbolFont{largesymbols}{OMX}{cmex}{m}{n}

\usepackage{hyperref}% add hypertext capabilities
\hypersetup{
    bookmarks=true,         % show bookmarks bar?
    unicode=false,          % non-Latin characters in Acrobatâs bookmarks
    pdftoolbar=true,        % show Acrobatâs toolbar?
    pdfmenubar=true,        % show Acrobatâs menu?
    pdffitwindow=false,     % window fit to page when opened
    pdfstartview={FitH},    % fits the width of the page to the window
    pdfsubject={Plasma fluid theory},   % subject of the document
    pdfproducer={GNU Make}, % producer of the document
    pdfkeywords={keyword1} {key2} {key3}, % list of keywords
    pdfnewwindow=true,      % links in new PDF window
    colorlinks=true,        % false: boxed links; true: colored links
    linkcolor=blue,      % color of internal links (change box color with linkbordercolor)
    citecolor=blue,         % color of links to bibliography
    filecolor=blue,      % color of file links
    urlcolor=blue           % color of external links
}
\usepackage{array}

\newcommand{\mhe}[1]{\bm{He}_{(#1)}}
\newcommand{\mh}[1]{\bm{H}_{(#1)}}

\newcommand{\mhh}[1]{\bm{\bar{H}}_{(#1)}}
\newcommand{\mg}[1]{\bm{\bar{G}}_{(#1)}}
\newcommand{\intR}[1]{\int_{\mathbb{R}^3}d\bm{#1}^3\;}

\newcommand{\revision}[1]{{#1}}

\makeatother

\begin{document}
\title{Exact collisional moments for plasma fluid theories}

\author{D. Pfefferl\'e}
\affiliation{Princeton Plasma Physics Laboratory, Princeton, New Jersey 08543, USA}
\author{E. Hirvijoki}
\affiliation{Princeton Plasma Physics Laboratory, Princeton, New Jersey 08543, USA}
\author{M. Lingam}
\affiliation{Department of Astrophysical Sciences, Princeton University, Princeton, NJ 08544, USA}
\affiliation{Harvard John A. Paulson School of Engineering and Applied Sciences, Harvard University, Cambridge, MA 02138, USA}
\affiliation{Harvard-Smithsonian Center for Astrophysics, The Institute for Theory and Computation, Cambridge, MA 02138, USA}

\begin{abstract}
  \revision{The velocity-space moments of the often troublesome
    nonlinear Landau collision operator are expressed exactly in terms
    of multi-index Hermite-polynomial moments of the distribution
    functions. The collisional moments are shown to be generated by
    derivatives of two well-known functions, namely the
    Rosenbluth-MacDonald-Judd-Trubnikov potentials for a Gaussian
    distribution. The resulting formula has a nonlinear dependency on
    the relative mean flow of the colliding species normalised to the
    root-mean-square of the corresponding thermal velocities, and a
    bilinear dependency on densities and higher-order velocity moments
    of the distribution functions, with no restriction on temperature,
    flow or mass ratio of the species. The result can be applied to
    both the classic transport theory of plasmas, that relies on the
    Chapman-Enskog method, as well as to deriving collisional fluid
    equations that follow Grad's moment approach. As an illustrative
    example, we provide the collisional ten-moment equations with
    exact conservation laws for momentum- and energy-transfer rate.}
\end{abstract}

\pacs{put Pacs here}%
 
\maketitle

\section{Introduction}
Fluid models have been widely employed in many fields of science,
ranging from astronomy and physics to biology and chemistry. The
fundamental principle, and motivation, behind fluid models is to
provide an effective macroscopic representation of the collective
behaviour arising from a large number of microscopic events. Thus, the
main advantage of fluid models is a reduction in complexity, while
still capturing the essential characteristics of the macroscopic
system.

Typically, the fluid equations are derived from a parent kinetic model
\cite{197Ichi,Villani2002} in which the particle dynamics is governed
by the equation
\begin{equation}\label{eq:kinetic-equation}
\frac{d f_{s}}{dt}=\sum_{s'}C_{ss'}[f_{s},f_{s'}].
\end{equation}
In the above formula, $f_s(t,\bm{x},\bm{v})$ denotes the phase-space
distribution function of species $s$, and $d/dt=\partial/\partial t +
\dot{\bm{x}}\cdot\nabla_{\bm{x}} + \dot{\bm{v}}\cdot\nabla_{\bm{v}}$
is the free-streaming Vlasov operator. On the RHS,
$C_{ss'}[f_{s},f_{s'}]$ is the (bilinear) collision operator between
the particle species $s$ and $s'$, thereby embodying the transition
from many-body dynamics to the dynamical evolution of a
single-particle distribution function.

The difficulty in constructing fluid models from kinetic theory arises
from the presence of the collision operator on the RHS of
(\ref{eq:kinetic-equation}). There exist two primary approaches for
addressing this issue: 1) the Chapman-Enskog
procedure~\cite{Chapman:1970} is a perturbation theory that relies
upon a small-parameter expansion in the Knudsen number of the kinetic
equation, and 2) Grad's procedure~\cite{Grad:1949_rarefied_gas_theory}
is a Galerkin projection based on the expansion of the distribution
functions in terms of orthogonal polynomials. The starting point in
both cases is a Maxwellian distribution function, corresponding to the
null space of the collision operator, \revision{while further
  refinements require evaluation of velocity moments of the collision
  operator.}

Recently, the fluid moments of the nonlinear Landau collision operator
were provided in a systematic and programmable
way~\cite{hirvijoki-2016}, and were demonstrated to be generated by
the gradients of three scalar valued integrals. However, closed-form
expressions for the integrals were not found. In the present work, we
improve upon previous findings \cite{hirvijoki-2016}, and provide this
time a closed, analytic form.
\revision{The derivation exploits some remarkable properties of the
  Hermite polynomials and the Maxwellian distribution function under
  convolution. The result is expressed as gradients of two well-known
  functions, namely the Rosenbluth-MacDonald-Judd-Trubnikov potentials
  \citep{RMJ1957,Trubnikov1958} for a Gaussian distribution, taken
  with respect to a dimensionless variable denoting the relative mean
  flow of the colliding species normalised to root-mean-square of the
  corresponding thermal velocities. By this fact, the result is
  manifestly Galilean invariant, which is a property that is usually
  not preserved under various approximations of the Landau collision
  operator. The formula is also bilinear with respect to species
  densities and the so-called second- or higher-order Hermite moments
  of the distribution functions.
  The procedure is valid regardless of the mass ratio, temperature or
  flow difference between species and, since the Hermite polynomials
  form a complete basis, it is exact. In other words, the knowledge of
  the Hermite moments of any distribution function is sufficient to
  provide the collisional moments of the nonlinear Landau operator
  exactly.

  Our representation is most convenient for the hierarchy of moment
  equations obtained via Grad's approach. Generating extended
  collisional fluid equations for plasmas to arbitrary order is then
  expected to be straight-forward with the help of computer algebra
  systems. Given the equivalence between Laguerre and contracted
  multi-index Hermite polynomials -- discussed in detail in appendix
  \ref{app:laguerre} -- a linearised version of our general formula
  can also be used within a Chapman-Enskog approach to recover
  Braginskii's transport coefficients, and possibly to extend the
  calculation to arbitrary species, flows and temperature
  differences.}

\revision{The paper begins in Sec.~\ref{sec:ce_discussion} with a
  thorough discussion on the similarities and differences between the
  Chapman-Enskog and Grad's approach. In particular, it is motivated
  why truncated distribution functions are required in both cases.}
Sec.~\ref{sec:hermite_expansion} recaps Grad's Hermite expansion
\revision{for square-integrable functions} and provides important
definitions and identities for further computations. After presenting
the necessary tools, the collisional moments are given explicit
expressions in Sec.~\ref{sec:landau_moments}. \revision{To consider
  how the result could be applied to the Chapman-Enskog theory, a
  correspondence between the Laguerre expansion -- typically used to
  solve the so-called correction equations -- and Grad's Hermite
  expansion of the distribution function is established in
  Sec.~\ref{sec:ce_hermite}.}  To illustrate how the resulting
formulae can be applied in the context of Grad's moment approach, the
collisional ten-moment fluid equations are derived in
Sec.~\ref{sec:ten-moment} and the nonlinear expressions for the
momentum- and energy-transfer rate are proven to exactly satisfy the
conservation laws. Sec.~\ref{sec:conclusion} concludes the work.
\section{Approaches to fluid theories}
\label{sec:ce_discussion}
\paragraph{Chapman-Enskog:} With the Chapman-Enskog approach,
near-Maxwellian corrections are derived through a hierarchy of linear
asymptotic equations, order by order. At each order, an integral
equation must be solved that involves the Vlasov operator acting on
the distribution function of the previous order on the LHS and the
collision operator acting linearly on the current order on the
RHS. Spatial gradients of the Maxwellian variables (temperature,
density and flow) thus become associated with the collisional moments
of higher-order corrections to the distribution
function. \revision{The parameters in front of the spatial gradients
  are collected as transport coefficients; for example at first order,
  the stress tensor is proportional to the strain tensor (or velocity
  gradient) via \emph{viscosity} and the heat flux to temperature
  gradient via \emph{conductivity}. For neutral fluids, the Euler
  equations, which originate from purely Maxwellian behaviour in
  velocity space, become at first-order the Navier-Stokes equations
  and at higher-order the Burnett and super-Burnett equations.

  For magnetised plasmas, the Chapman-Enskog procedure is a more
  complex multi-parameter perturbation theory due to the dominance of
  the Lorentz force, the long-range Coulomb interactions and the
  presence of multiple species. The first-order solution is known as
  the Braginskii equations~\cite{Braginskii:1965} and is commonly used
  to describe (classical) transport in magnetised plasmas where the
  mean-free-path is comparable to the Larmor radius but shorter than
  characteristic gradient scale lengths. Different orderings and
  second-order solutions have been attempted
  \cite{mikhailosvkii-tsypin-1971,catto-simakov-2004}. In order to
  make the problem tractable, the Landau collision operator
  \cite{Landau_1937} between different species must be approximated
  and/or split into simpler forms, especially in the case where the
  species flows and temperatures are different \cite{Ram07}. In this
  regard, the Maxwellian solution and thus the starting point for the
  perturbative treatment of the kinetic equation is valid only in the
  limit of vanishing mass ratio between electron and ion species, in
  which case the electron-ion collision operator is approximated by
  the Lorentz collision operator.

  Solving the integral equation at the next order is far from
  trivial. By virtue of the self-adjoint properties of the linearised
  collision operator, it is typically reformulated as a variational
  problem \cite{robinson-bernstein-1962}, where a functional of the
  solution is maximised by varying the coefficients of a polynomial
  expansion of the distribution function. Depending on the choice of
  polynomial basis (Sonine, Laguerre or Legendre), the transport
  coefficients obtained converge rapidly to their physical values as a
  function of the number of terms in the
  series~\cite{hinton-hazeltine-1976}. The Chapman-Enskog procedure,
  applied to plasmas or neutral fluids, is thus formally solved using
  a truncated polynomial expansion of the distribution function.}

\paragraph{Grad: } Grad's method relies upon solving the
kinetic equation indirectly by projecting it onto a set of orthogonal
polynomials, namely the multi-index Hermite polynomials. \revision{The
  Hermite basis naturally arises from the Gaussian measure, i.e. the
  null-space solution of the collision operator}. The procedure yields
a weak solution to the kinetic equation, the same way finite-element
methods \cite{brenner2007mathematical} are constructed, or
\emph{observables} are obtained as expectation values of the
Schr\"odinger equation in quantum mechanics
\citep{Wein12}. \revision{The result is an infinite hierarchy of
  dynamical equations for the projection coefficients. The projection
  coefficients correspond to moments of the distribution function and
  are evolved in time by these equations from known initial
  conditions. The evolution of all moments is equivalent to the
  evolution of the distribution function, in the sense that the
  complete set of moment equations is a spectral representation of the
  kinetic equation. In its infinite form, this method remains a
  microscopic description of the fluid at hand \cite[Chapter
  4.5]{balescu-1988}.}

A closed set of fluid equations is obtained by setting, above a given
order, all expansion coefficients to zero, thereby restricting the
solution of the distribution function to a finite dimensional vector
space. \revision{The justification for a specific truncation depends
  on the studied flow and is done \emph{a posteriori} by verifying
  certain realisability conditions
  \cite[Eq.(3.24)]{levermore-1996}. We note that these realisability
  conditions also apply to the Chapman-Enskog approach.}
Orthogonality among Hermite polynomials guarantees that all the fluid
contributions of a given order have been properly captured. The
projection of the collision operator does not yield spatial gradients,
and therefore the collisional moments are not immediately related to
orthodox transport coefficients, as in the Chapman-Enskog approach.

\revision{In the limit of vanishing mean-free-path, one can apply a
  Chapman-Enskog analysis to Grad's hierarchy of moment equations and
  show under certain assumptions on timescales and flow that they
  faithfully reduce to the Navier-Stokes equations
  \cite{Grad:1949_rarefied_gas_theory}. The effect of the truncation
  is mainly to underestimate the coefficient of viscosity, as proven
  for generic collision operators in neutral fluids by Levermore
  \cite[Eq.(5.30)]{levermore-1996}. Levermore also shows that there is
  a natural way of ordering corrections to transport coefficients
  involving higher-order moments; a BGK analysis of the collision
  operator in the limit of zero mean-free-path reveals that the
  associated relaxation rates are increasingly stronger
  \cite[Eq.(6.28)]{levermore-1996}. Grad's moment equations, although
  limited in the accuracy of asymptotic transport coefficients, have
  the advantage over Chapman-Enskog transport equations of retaining
  more kinetic features and being valid beyond Maxwellian behaviour,
  as discussed in \cite[Appendix 4]{Grad:1949_rarefied_gas_theory},
  thereby embedding classical transport theory as a special case of
  its realisable flows}. Grad's method is thus often applied to the
investigation of the nonlinear properties of fluids and the formation
of shocks in rarefied gases.

\revision{In plasma physics, Grad's moment equations are not used
  nearly as much as Braginskii's, even though the trial solution to
  the first order Chapman-Enskog correction equation written with
  Laguerre polynomials corresponds identically to a contracted
  multi-index Hermite polynomial expansion. This means that the trial
  solution with only the first ($N=1$) Laguerre polynomial coincides
  with Grad's 13-moment expansion and the trial solution with $N=2$
  truncation matches Grad's 21-moment and so on. This correspondence
  was noticed, e.g., by Balescu \cite[chapter 4]{balescu-1988} and is
  revisited in section \ref{sec:ce_hermite} for the expansion
  coefficients in terms of which the collisional moments of the Landau
  operator are expressed.}
% can be proven explicitly, as shown in our appendix
% \ref{app:laguerre} and \cite[Appendix G1.4]{balescu-1988}.} The
% reluctance to apply Grad's approach in plasma physics possibly stems
% from the algebraic difficulty of computing the moments of the Landau
% collision operator while retaining the result in terms of fluid
% quantities~\cite{JH09}.
%
\section{Hermite expansion of square-integrable functions}
\label{sec:hermite_expansion}
Inspired by the seminal work by Grad on the asymptotic theory of the
Boltzmann equation \cite{Grad:1949_rarefied_gas_theory}, we
\revision{consider the spectral expansion of square-integrable
  functions} in terms of multi-index Hermite polynomials. While
various definitions exist, we will employ, from the review by
Holmquist \cite{Holmquist:1996}, the so-called \emph{covariant Hermite
  polynomials}
\begin{equation}
  \label{eq:covariant_gen}
  \mg{k}(\bm{x}-\bm{\mu};\sigma^2) = \frac{1}{\mathcal{N}_{\sigma^2}(\bm{x}-\bm{\mu})}(-\nabla_{\bm{x}})^{(k)}   \mathcal{N}_{\sigma^2}(\bm{x}-\bm{\mu}), 
\end{equation}
as well as the so-called \emph{contravariant Hermite polynomials}
\begin{align}\label{eq:hermite_def}
\mhh{k}(\bm{x}-\bm{\mu};\sigma^2) = \frac{1}{\mathcal{N}_{\sigma^2}(\bm{x}-\bm{\mu})}(-\sigma^2\nabla_{\bm{x}})^{(k)} \mathcal{N}_{\sigma^2}(\bm{x}-\bm{\mu}) = \sigma^{2k}\mg{k}(\bm{x}-\bm{\mu};\sigma^2),
\end{align}
both generated by the three-dimensional Normal distribution
\begin{equation}\label{eq:normal_distro}
\mathcal{N}_{\sigma^2}(\bm{x}-\bm{\mu})\equiv \frac{e^{-(\bm{x}-\bm{\mu})^2/2\sigma^2}}{(2\pi)^{3/2}\sigma^3}.
\end{equation}
Regarding our notation for outer products of vectors (such as
consecutive application of the gradient operator), it is implied
throughout this document that
\begin{equation}
\nabla^{(k)}\equiv\underbrace{\nabla\otimes\dots\otimes\nabla}_{k \textrm{ terms}}, \qquad \bm{x}^{(k)}\equiv\underbrace{\bm{x}\otimes\dots\otimes\bm{x}}_{k \textrm{ terms}},
\end{equation}
whereas for tensors (such as the Hermite polynomials), the notation
refers to the rank of the multi-indexing according to
\begin{equation}
\mg{k}(\bm{x})=\mg{k}^{k_1\dots k_k}(\bm{x}) \quad \textrm{where}\quad k_i\in\{1,2,3\}.
\end{equation}

One important property of the Hermite polynomials is that they are
orthogonal to each other with respect to the Gaussian measure,
\begin{align}\label{eq:orthonormality}
  \int_{\mathbb{R}^3} d\bm{y} \mhh{i}(\bm{y};\sigma^2) \mg{j}(\bm{y};\sigma^2)\mathcal{N}_{\sigma^2}(\bm{y}) %\overset{(\ref{eq:grad_gaussian})}{=}\nabla_{\bm{x}}^{(j)} \int_{\mathbb{R}^3} d\bm{y} \mhh{i}(\bm{y};\sigma^2)\mathcal{N}_{\sigma^2}(\bm{y}-\bm{x})\Big|_{\bm{x}=\bm{0}}\nonumber\\
\overset{(\ref{eq:grad_gaussian},\ref{eq:weierstrass_hermite})}{=}\nabla_{\bm{x}}^{(j)}\bm{x}^{(i)}\Big|_{\bm{x}=\bm{0}} = \delta^{(i)}_{[(j)]},
% \int_{\mathbb{R}^3} d\bm{y} \mhh{i}(\bm{y};\sigma^2) \mg{j}(\bm{y};\sigma^2)\mathcal{N}_{\sigma^2}(\bm{y})
% &\overset{(\ref{eq:grad_gaussian},\ref{eq:weierstrass_hermite})}{=}\nabla^j\bm{x}^i\Big|_{\bm{x}=\bm{0}} \equiv \delta^{(i)}_{[(j)]},
\end{align}
where $[(j)] = [j_1 \cdots j_j] = \sum_\pi \pi(j_1)\cdots \pi(j_j)$ is
the sum over all permutations of $j$ indices and
$\delta^{(i)}_{(j)} = \delta^{i_1}_{j_1}\dots\delta^{i_i}_{j_i}$. This
orthogonality property allows for a convenient expansion \revision{of any square-integrable function}, such as the distribution function $f_s$ for species $s$, according to
\begin{align}
\label{eq:charlier_expansion}
  \frac{f_s(\bm{v})}{n_s}  = \mathcal{N}_{\sigma_s^2}(\bm{v}-\bm{V}_s) \sum_{i=0}^\infty \frac{1}{i!}\bm{c}_{s(i)}\mg{i}(\bm{v}-\bm{V}_s;\sigma_s^2)%\nonumber\\
\overset{(\ref{eq:covariant_gen})}{=}\sum_{i=0}^\infty\frac{\bm{c}_{s(i)}}{i!}\nabla_{\bm{V}_s}^{(i)}\mathcal{N}_{\sigma^2_s}(\bm{v}-\bm{V}_s),
\end{align}
where the (symmetric) expansion coefficients are the so-called
\emph{Hermite-moments} of the distribution function
\begin{align}\label{eq:distro_hermites}
\bm{c}_{s(j)}\equiv \int_{\mathbb{R}^3} d\bm{v} \frac{f_s(\bm{v})}{n_s}\mhh{j}(\bm{v}-\bm{V}_s;\sigma_s^2). 
%&= \sum_{i=0}^\infty \frac{\bm{c}_{(i)}}{i!}\intR{v} \mg{i}(\bm{v}-\bm{V};\sigma^2)\mhh{j}(\bm{v}-\bm{V};\sigma^2) \mathcal{N}_{\sigma^2}(\bm{v}-\bm{V}). 
%\overset{(\ref{eq:orthonormality})}{=} \sum_{i=0}^\infty \frac{1}{i!}\bm{c}_{\bm{i}} \delta^{\bm{i}}_{[\bm{j}]} = \frac{\bm{c}_{[j]}}{j!}\equiv\bm{c}_{(j)}.
\end{align}
As per the Einstein summation convention on repeated indices $(i)$,
the tensors $\mg{i}(\bm{x};\sigma^2)$ in the expansion are fully
contracted with the tensors $\bm{c}_{(i)}$. 

Written in the form (\ref{eq:charlier_expansion}), the distribution
function of each species is automatically normalised to the species'
density and thus $c_{s(0)} = 1$. The mean velocity, $\bm{V}_s$, and the
variance of each species (half thermal velocity squared),
$\sigma_s^2=\tfrac{1}{2} v_{\text{th},s}^2=T_s/m_s$, is contained in
the Gaussian envelope, so that $\bm{c}_{s(1)} = \bm{0}$ and
$\bm{c}_{s(2)} = (\bm{P}_s - p_s\bm{I})/n_s m_s =
\sigma_s^2(\bm{P}_s/p_s - \bm{I})$ represents the trace-less pressure
tensor measuring the degree of anisotropy and off-diagonal features,
where $m_s$, $n_s$ and $T_s$ are the species' mass, density and
temperature respectively, $p_s=n_s T_s = \frac{1}{3}\text{tr}\bm{P}_s$
is the isotropic pressure and $\bm{P}_s \equiv m_s \intR{v}
(\bm{v}-\bm{V}_s)(\bm{v}-\bm{V}_s) f_s$ is the pressure tensor. %
\revision{Third-order tensorial moments of the distribution function
  are captured by the tensor
  $\bm{c}_{s(3)} = \intR{v}
  (\bm{v}-\bm{V}_s)(\bm{v}-\bm{V}_s)(\bm{v}-\bm{V}_s) f_s/n_s$ which,
  when maximally contracted, represents the heat-flux,
  $q^i_s = m_sn_s c_{s(3)}^{ikk}= m_s \intR{v}
  |\bm{v}-\bm{V}_s|^2(\bm{v}-\bm{V}_s)f_s$.}

The temporal and spatial dependence of the kinetic distribution
$f_s(\bm{v})$ has been omitted for convenience, but it is understood
that the coefficients $n_s$, $\bm{V}_s$, $T_s$ and $\bm{c}_{s(i)}$
(i.e. the fluid variables) vary with respect to time and spatial
coordinates. \revision{It is also important to note that the definition for the
  projection coefficients (\ref{eq:distro_hermites}) is independent of
  how the distribution function is presented.}

% In standard fluid theory, one usually considers the monomial
% $\bm{v}^{(n)}$ moments of (\ref{eq:kinetic-equation}), since they are
% straightforward to evaluate for the free-streaming Vlasov
% operator. For the collision operator, which is local in velocity
% space, the moments with respect to the Hermite polynomials
% $\mhh{n}(\bm{v}-\bm{V}_s;\sigma_s^2)$ turn out to be more practical
% instead. Since both are polynomials, one can easily perform a change
% of basis via
% \begin{equation}\label{eq:monomials-to-hermites}
% \bm{v}^{(n)}=\sum_{l=0}^n\bm{b}^{(n)}_{s(l)}\mhh{l}(\bm{v}-\bm{V}_s;\sigma_s^2),
% \end{equation}
% and invoke the orthogonality relation, while noting that $\bm{v}^{(n)} = \mhh{n}(\bm{v};0)$,
%  to compute the coefficients
% \begin{equation}\label{eq:monomials-to-hermites-coeff}
% \bm{b}^{(n)}_{s(l)}=\intR{v} \bm{v}^{(n)}\mg{l}(\bm{v}-\bm{V}_s;\sigma_s^2)\mathcal{N}_{\sigma_s^2}(\bm{v}-\bm{V}_s).
% \end{equation}

Before computing the Hermite-moments of the collision operator, one
more identity is needed, namely the directional derivative of a given
Hermite polynomial. For any vector $\bm{J}$, one
has~\citep[Eq.(6.2)]{Holmquist:1996}
\begin{equation}
\label{eq:hermite_gradient}
  \bm{J}\cdot\nabla_{\bm{v}} \mhh{k}(\bm{v}-\bm{V};\sigma^2) 
%= J^i\partial_i  \mhh{k}^{k_1\cdots k_k}(\bm{v}-\bm{V};\sigma^2) 
  = \frac{1}{(k-1)!} J^{[k_1}\mhh{k-1}^{k_2\cdots k_{k}]}(\bm{v}-\bm{V};\sigma^2) = k \text{ Sym}[ \bm{J} \mhh{k-1}(\bm{v}-\bm{V};\sigma^2)]
\end{equation}
where
$\text{Sym}\bm{A}^{a_1\cdots a_k} = \bm{A}^{[a_1 \cdots a_k]}/k!$ is
the symmetrization of a tensor. This result appears in Grad's original
work for dimensionless Hermite
polynomials~\citep[Eq.(17)]{Grad:1949_polynomials}.
\section{Landau Collision operator and Hermite-moments}
\label{sec:landau_moments}
In warm plasmas, collisions are dominated by continuous small-angle
Coulomb scattering. The appropriate operator to describe the
collective effect of these events was derived by
Landau~\cite{Landau_1937}, and can be expressed as a velocity-space
divergence of a collisional velocity-space flux defined by
\begin{equation}
\label{eq:collision_operator}
C_{ss'}[f_s,f_{s'}](\bm{v}) \equiv -\frac{c_{ss'}}{m_s}\nabla_{\bm{v}}\cdot\bm{J}_{ss'}[f_s,f_{s'}](\bm{v}).
\end{equation}
Here, $c_{ss'}=\ln\Lambda (e_se_{s'})^2/4\pi\varepsilon_0^2$,
$\ln\Lambda$ denotes the Coulomb Logarithm, and $e_s$ is the species
charge. The collisional velocity-space flux $\bm{J}_{ss'}$ can be
represented in its original integral form, or through the so-called
Rosenbluth-MacDonald-Judd-Trubnikov potential
functions~\cite{Trubnikov1958,RMJ1957} according to
\begin{equation}
\bm{J}_{ss'}[f_s,f_{s'}](\bm{v}) \equiv  \mu_{ss'}(\nabla_{\bm{v}}\phi_{s'}) f_s
-m_s^{-1}\nabla_{\bm{v}}\cdot\left[(\nabla_{\bm{v}}\nabla_{\bm{v}}\psi_{s'}) f_s\right],
\end{equation}
where $\mu_{ss'}=1/m_s+1/m_{s'}$ and the potential functions,
$\phi_s(\bm{v})$, and $\psi_s(\bm{v})$, are defined through
\begin{align}
\phi_s(\bm{v})&\equiv \int_{\mathbb{R}^3} d\bm{v}' f_s(\bm{v}')\,\lvert \bm{v}-\bm{v}'\rvert^{-1}, 
% = (f_s * |\cdot|^{-1})(\bm{v})
&
\psi_s(\bm{v})&\equiv \frac{1}{2}\int_{\mathbb{R}^3} d\bm{v}' f_s(\bm{v}')\,\lvert \bm{v}-\bm{v}'\rvert. 
% = \frac{1}{2}(f_s * |\cdot|)(\bm{v}).
\end{align}
%where $*$ denotes convolution. 
We observe that $\nabla_{\bm{v}}\cdot\nabla_{\bm{v}} \psi_s = \phi_s$ and $\nabla_{\bm{v}}\cdot\nabla_{\bm{v}}\phi_s = -4\pi f_s$.

%\subsection{Moments of the collision operator}
%
To derive collisional fluid equations based on Grad's expansion
(\ref{eq:charlier_expansion}), we consider the Hermite-moments of the
collision operator
\begin{align}
\bm{C}_{ss'(k+1)}&\equiv m_s \int_{\mathbb{R}^3} d\bm{v} \mhh{k+1}(\bm{v}-\bm{V}_s;\sigma_s^2) C_{ss'}(\bm{v}) \nonumber\\
% \overset{(\ref{eq:collision_operator})}{=} c_{ss'}\int_{\mathbb{R}^3} d\bm{v}\bm{J}_{ss'}(\bm{v})\cdot \nabla_{\bm{v}} \mhh{k+1}(\bm{v}-\bm{V}_s;\sigma_s^2) \\
% &\overset{(\ref{eq:hermite_gradient})}{=} c_{ss'}(k+1) \text{Sym}\int_{\mathbb{R}^3} d\bm{v}\bm{J}_{ss'}(\bm{v})\mhh{k}(\bm{v}-\bm{V}_s;\sigma_s^2)\\ &
& \equiv \bar{c}_{ss'}(k+1)\text{Sym}\left[\mu_{ss'}\bm{R}_{ss'(k+1)} + \frac{k}{m_s}\bm{D}_{ss'(k+1)}\right],
\end{align}
where $\bar{c}_{ss'}=n_sn_{s'}c_{ss'}$ and the integral has been split, after first integrating by parts
and then using identity~(\ref{eq:hermite_gradient}), into drag- and diffusion-related terms
\begin{align}
\label{eq:R-def}
\bm{R}_{ss'(k+1)} &= \frac{1}{n_sn_{s'}}\int_{\mathbb{R}^3} d\bm{v}\, (\nabla_{\bm{v}} \phi_{s'})\, f_s\mhh{k}(\bm{v}-\bm{V}_s;\sigma_s^2)\\
\label{eq:D-def}
\bm{D}_{ss'(k+1)} 
%= \int_{\mathbb{R}^3}\left[(\nabla_{\bm{v}}\nabla_{\bm{v}}\psi_{s'}) f_s\right] \cdot \nabla_{\bm{v}}\mhh{k}(\bm{v}-\bm{V}_s;\sigma_s^2)
&= \frac{1}{n_sn_{s'}}\int_{\mathbb{R}^3} d\bm{v}\,(\nabla_{\bm{v}}\nabla_{\bm{v}}\psi_{s'})\, f_s \mhh{k-1}(\bm{v}-\bm{V}_s;\sigma_s^2)
\end{align}
One may already notice that $C_{ss'(0)}=0$ and
$\bm{D}_{ss'(1)}=\bm{0}$ such that
$\bm{C}_{ss'(1)} = \bar{c}_{ss'}\mu_{ss'}\bm{R}_{ss'(1)}\equiv\bm{F}_{ss'}$
represents the collisional momentum transfer rate between species $s$
and $s'$.

%\paragraph{Drag term:}
To proceed, the velocity gradients of the potential functions are
manipulated in order to extract derivatives with respect to the mean
velocity $\bm{V}_{s'}$ according to
\begin{align}
\frac{1}{n_{s'}}\nabla_{\bm{v}} \phi_{s'}(\bm{v}) 
%= \int_{\mathbb{R}^3} d\bm{v}' f_{s'}(\bm{v}')\nabla_{\bm{v}}|\bm{v}-\bm{v}'|^{-1} 
%= \int_{\mathbb{R}^3} d\bm{v}'\frac{\nabla_{\bm{v}'}f_{s'}(\bm{v}')}{|\bm{v}-\bm{v}'|}
& =-\nabla_{\bm{V}_{s'}}\sum_{j=0}^\infty\frac{c_{s'(j)}}{j!}\nabla_{\bm{V}_{s'}}^{(j)}\int_{\mathbb{R}^3} d\bm{v}'\frac{\mathcal{N}_{\sigma^2_{s'}}(\bm{v}'-\bm{V}_{s'})}{|\bm{v}-\bm{v}'|}, \\
\frac{1}{n_{s'}}\nabla_{\bm{v}}\nabla_{\bm{v}} \psi_{s'}(\bm{v})
%&= \frac{1}{2}\int_{\mathbb{R}^3} d\bm{v}' f_{s'}(\bm{v}')\nabla_{\bm{v}}\nabla_{\bm{v}}|\bm{v}-\bm{v}'|
%&= \int_{\mathbb{R}^3} d\bm{v}'\, \nabla_{\bm{v}'}f_{s'}(\bm{v}')\nabla_{\bm{v}}|\bm{v}-\bm{v}'|\\
%=\frac{1}{2} \int_{\mathbb{R}^3} d\bm{v}'\, \nabla_{\bm{v}'}\nabla_{\bm{v}'}f_{s'}(\bm{v}')|\bm{v}-\bm{v}'|
& = \frac{1}{2}\nabla_{\bm{V}_{s'}}\nabla_{\bm{V}_{s'}}\sum_{j=0}^\infty\frac{\bm{c}_{s'(j)}}{j!}\nabla_{\bm{V}_{s'}}^{(j)}\int_{\mathbb{R}^3} d\bm{v}'\mathcal{N}_{\sigma^2_{s'}}(\bm{v}'-\bm{V}_{s'})|\bm{v}-\bm{v}'|.
\end{align}
Next, the products of two Hermite polynomials in
$f_s(\bm{v})\mhh{k}(\bm{v}-\bm{V}_s;\sigma_s^2)$ are expressed as a
series of single Hermite polynomials (linearisation) using several
identities derived in Appendix~\ref{sec:hermite_properties} so to
extract derivatives with respect to the mean velocity $\bm{V}_s$
according to
\begin{align}
\frac{1}{n_s}f_s(\bm{v})\mhh{k}(\bm{v}-\bm{V}_s;\sigma_s^2)
% &\overset{(\ref{eq:covariant_hermites_def})}{=}\mathcal{N}_{\sigma^2_s}(\bm{v}-\bm{V}_s)\sum_{i=0}^{\infty}\frac{\bm{c}_{s(i)}}{i!\sigma_s^{2i}}\mhh{i}(\bm{v}-\bm{V}_s;\sigma_s^2)\mhh{k}(\bm{v}-\bm{V}_s;\sigma_s^2)\nonumber\\
% &\overset{(\ref{eq:linearization})}{=}\mathcal{N}_{\sigma^2_s}(\bm{v}-\bm{V}_s)\sum_{i=0}^{\infty}\sum_{l=0}^{i+k}\frac{\bm{c}_{s(i)}}{i!}\frac{\bm{a}^{(l)}_{(i)(k)}}{\sigma_s^{2i}}\mg{l}(\bm{v}-\bm{V}_s;\sigma_s^2)\nonumber\\
&\overset{(\ref{eq:covariant_hermites_def}),(\ref{eq:grad_gaussian}),(\ref{eq:linearization}), (\ref{eq:neutral_coeff})}{=}\sum_{i=0}^{\infty}\sum_{l=0}^{i+k}\frac{\bm{c}_{s(i)}}{i!}\sigma_s^{k+l-i}\bm{\bar{a}}^{(l)}_{(i)(k)}\nabla_{\bm{V}_s}^{(l)} \mathcal{N}_{\sigma^2_s}(\bm{v}-\bm{V}_s)
\label{eq:fhk}.
\end{align}
The expression for the so-called linearisation coefficient
$\bm{\bar{a}}^{(l)}_{(i)(j)}$, derived explicitly in
Appendix~\ref{sec:linearisation}, is
\begin{align}
\bm{\bar{a}}^{(l)}_{(i)(j)} &= \frac{1}{l!}\nabla_{\bm{x}}^{(i)}\nabla_{\bm{y}}^{(j)}\nabla_{\bm{z}}^{(l)} \left[e^{\bm{x}\cdot\bm{y} + \bm{y}\cdot\bm{z} + \bm{x}\cdot\bm{z}}\right]_{\bm{x}=\bm{0},\bm{y}=\bm{0},\bm{z}=\bm{0}}.
\end{align}

Since the gradients with respect to $\bm{V}_s$ and $\bm{V}_{s'}$ can
be brought out of the integrals~(\ref{eq:R-def}) and~(\ref{eq:D-def}), one
is only left with the task of evaluating the following integrals
\begin{align}
\iint_{\mathbb{R}^3} d\bm{v} d\bm{v}'\frac{\mathcal{N}_{\sigma_{s'}^2}(\bm{v}'-\bm{V}_{s'})\mathcal{N}_{\sigma_s^2}(\bm{v}-\bm{V}_s)}{|\bm{v}-\bm{v}'|}&=\frac{1}{\sqrt{2}\Sigma_{ss'}}\Phi\left(\frac{|\bm{U}_{ss'}|}{\sqrt{2}\Sigma_{ss'}}\right),\label{eq:convo_drag}\\
\frac{1}{2}\iint_{\mathbb{R}^3} d\bm{v} d\bm{v}'\, \mathcal{N}_{\sigma_{s'}^2}(\bm{v}'-\bm{V}_{s'})\mathcal{N}_{\sigma_s^2}(\bm{v}-\bm{V}_s)|\bm{v}-\bm{v}'|&=\sqrt{2}\Sigma_{ss'}\Psi\left(\frac{|\bm{U}_{ss'}|}{\sqrt{2}\Sigma_{ss'}}\right),\label{eq:convo_diffusion}
\end{align}
where $\bm{U}_{ss'}=\bm{V}_s-\bm{V}_{s'}$,
$\Sigma_{ss'}^2=\sigma_s^2+\sigma_{s'}^2=\frac{1}{2}(v^2_{\text{th},s}+v^2_{\text{th},s'})$,
and the functions $\Phi(z)$ and $\Psi(z)$ are nothing but the
Rosenbluth-MacDonald-Judd-Trubnikov potentials for a Normal
distribution~${\cal N}_{1/2}(z)$, defined according to
\begin{align}
  \Phi(z) &= \int_{\mathbb{R}^3}\frac{d\bm{x}}{|\bm{x}|}\frac{e^{-(\bm{x}-\bm{z})^2}}{\pi^{3/2}}=\frac{\text{erf}(z)}{z},\\
  \Psi(z) &= \frac{1}{2}\int_{\mathbb{R}^3}d\bm{x}|\bm{x}|\frac{e^{-(\bm{x}-\bm{z})^2}}{\pi^{3/2}}=
\left(z + \frac{1}{2z}\right)\text{erf}(z) + \frac{e^{-z^2}}{\sqrt{\pi}}.
\end{align}
Intermediate steps to yield
(\ref{eq:convo_drag}) and (\ref{eq:convo_diffusion}) rely on the fact that
the convolution of two Normal distribution results in a Normal
distribution with the sum of the variances, as seen from equation
(\ref{eq:convolution_gaussians}).

The final task is to define a dimensionless parameter
$\bm{\Delta}_{ss'}=-\bm{\Delta}_{s's} =
\bm{U}_{ss'}/\sqrt{2}\Sigma_{ss'}$, and to transform gradients with
respect to $\bm{V}_s$ and $\bm{V}_{s'}$ into gradients with respect to
$\bm{\Delta}_{ss'}$. In physical terms, $\bm{\Delta}_{ss'}$ is the
ratio of the relative mean flow and (total) thermal velocities. As a
result, the drag- and diffusion-related term are expressed as
\begin{align}
\label{eq:drag_term}
\bm{R}_{ss'(k+1)}
&=\nabla_{\bm{\Delta}_{ss'}}\sum_{i,j=0}^\infty\sum_{l=0}^{i+k} \frac{(-1)^j\sigma_s^{k+l-i}}{(\sqrt{2}\Sigma_{ss'})^{l+j+2}}\frac{\bm{c}_{s(i)}}{i!}\frac{\bm{c}_{s'(j)}}{j!}\bm{\bar{a}}^{(l)}_{(i)(k)}\nabla_{\bm{\Delta}_{ss'}}^{(l)}\nabla_{\bm{\Delta}_{ss'}}^{(j)}
 \Phi(\Delta_{ss'}), 
\\
\label{eq:diffusion_term}
\bm{D}_{ss'(k+1)}
&=\nabla_{\bm{\Delta}_{ss'}}\nabla_{\bm{\Delta}_{ss'}}\sum_{i,j=0}^\infty\sum_{l=0}^{i+k-1} \frac{(-1)^j\sigma_s^{k-1+l-i}}{(\sqrt{2}\Sigma_{ss'})^{l+j+1}}\frac{\bm{c}_{s(i)}}{i!}\frac{\bm{c}_{s'(j)}}{j!}\bm{\bar{a}}^{(l)}_{(i)(k-1)}\nabla_{\bm{\Delta}_{ss'}}^{(l)}\nabla_{\bm{\Delta}_{ss'}}^{(j)}
\Psi(\Delta_{ss'}), 
\end{align}
where $\Delta_{ss'} = |\bm{\Delta}_{ss'}|$. 
\revision{The collisional moments of the nonlinear Landau operator can
  thus be computed exactly, given the knowledge of the projection
  coefficients (\ref{eq:distro_hermites}) of the species distribution
  functions. Since the Hermite polynomials form a complete basis, the
  result is independent of how the distribution function is
  presented. It is directly applicable to Grad's expansion but can be
  accommodated to other polynomials (such as Laguerre) in the context
  of the Chapman-Enskog procedure, as discussed in section
  \ref{sec:ce_hermite}.

  The convergence of the bilinear series depends on the ratio between
  the Hermite-moments $\bm{c}_{s(i)}$ and the $i$-th power of the
  (total) variance $\Sigma_{ss'} = \sqrt{v_{th,s}^2 + v_{th,s'}^2}$,
  multiplied by the $i$-th gradient of the special functions $\Phi$
  and $\Psi$. The latter are functions of the dimensionless parameter
  $\Delta_{ss'}$ which is a small parameter in most physical cases
  \cite{LHP16}; for ion-electron plasmas,
  $\Delta_{ei} = \sqrt{m_e J^2/2ne^2p_e}\sim (d_e/L)\beta_e^{-1/2}\ll
  1$ on dimensional grounds (where $d_e/L$ is the normalised electron
  skin depth and $\beta_e$ the electron plasma beta). A Taylor
  expansion of the special functions around zero,
  \begin{align}
    \Phi(\Delta_{ei}) &= \frac{2}{\sqrt{\pi}} \sum_{n=0}^\infty \frac{(-1)^n \Delta_{ei}^{2n}}{n!(2n+1)},  &
    \Psi(\Delta_{ei}) &= \frac{2}{\sqrt{\pi}}\left[1 - \sum_{n=1}^\infty \frac{(-1)^n \Delta_{ei}^{2n}}{n!(2n+1)(2n-1)}\right],
  \end{align}
  confirms that applying any number of derivatives on these
  alternating fast-decaying series does not give rise to singularities
  nor does it affect the convergence of the collisional moments. The
  collisional momentum transfer rate is provided at lowest order and
  studied in more detail in the companion paper \cite{LHP16}.}

\section{Chapman-Enskog compatible Hermite expansion}
\label{sec:ce_hermite}
\revision{In order to solve the linear integral equation in the
  Chapman-Enskog approach (or the so-called Spitzer problem), the
  total distribution function is expressed as $f/n={\cal
    N}_{\sigma^2}(\bm{w})[1+\chi(\bm{w})]$, where the random velocity
  $\bm{w}=\bm{v}-\bm{V}$ is adopted for convenience. Considering the
  tensorial and vectorial invariance of the first-order correction
  equations, the term $\chi(\bm{w})$ is generically of the form
\begin{align}
\chi(\bm{w})=\sum_{n=2}^{\infty}a_{n}L_n^{1/2}\left(\frac{w^2}{2\sigma^2}\right)+\bm{w}\cdot\sum_{n=1}^{\infty}\bm{b}_{n}L_n^{3/2}\left(\frac{w^2}{2\sigma^2}\right)+\left(\bm{w}\bm{w}-w^2\frac{\bm{I}}{3}\right):\sum_{n=0}^{\infty}\mathbf{d}_{n}L_n^{5/2}\left(\frac{w^2}{2\sigma^2}\right),
\end{align}
where most notably, the first vector valued expansion coefficient
corresponds to heat-flux $\bm{b}_{1}=-m\bm{q}/5pT$ and the first
tensor valued expansion coefficient corresponds to viscosity
$\mathbf{d}_{0}=m(\bm{P}-p\bm{I})/2pT$, and is therefore traceless
$\text{tr}(\mathbf{d}_{0})=0$.
The linear integral equation in the Chapman-Enskog theory is then
efficiently converted into a linear algebraic equations for the
scalar, vector, and tensor coefficients $a_{n}$, $\bm{b}_{n}$, and
$\mathbf{d}_{n}$ \cite{robinson-bernstein-1962,Braginskii:1965}. Using
the results from Appendix \ref{app:laguerre}, an equivalent expression
for $\chi(\bm{w})$ can be given in terms of the irreducible Hermite
polynomials of scalar, vector and two-rank tensor kind as
\begin{align}
\chi(\bm{w})=&\sum_{n=2}^{\infty}\left(\frac{a_n}{N_n}-\frac{1}{3}\frac{\text{tr}(\mathbf{d}_{n-1})}{ N_{n-1}}\right)h_{|2n|}(w;\sigma^2)
+\sum_{n=1}^{\infty}\frac{\bm{b}_{n}}{N_n}\cdot \bm{h}_{|2n|+(1)}(\bm{w};\sigma^2)
+\sum_{n=0}^{\infty}\frac{\mathbf{d}_{n}}{N_n}:\bm{h}_{|2n|+(2)}(\bm{w};\sigma^2)
\end{align}
where $N_n = (-1)^n2^n \sigma^{2n} n!$ is the conversion factor
between the Laguerre and Hermite basis.
% and one can further observe that 
% \begin{align}
% \left(f_{n}^{-1}a_{n}-N_{n-1}^{-1}\text{tr}(\mathbf{d}_{n-1})/3\right)h_{|2n|}(w;\sigma^2)&=\sigma^{2(2n)}\left(f_{n}^{-1}a_{n}-N_{n-1}^{-1}\text{tr}(\mathbf{d}_{n-1})/3\right)\left(\delta_{i_1
%     i_2}\cdots\delta_{i_{2n-1}i_{2n}}\right) \mg{2n}^{i_1\ldots i_{2n}}(\bm{w};\sigma^2)\\
% N_n^{-1}\bm{b}_{n}\cdot \bm{h}_{|2n|+(1)}(\bm{w};\sigma^2)&=\sigma^{2(2n+1)}N_n^{-1}\bm{b}^j_{n} (\delta_{i_1
%     i_2}\cdots\delta_{i_{2n-1}i_{2n}})\mg{2n+1}^{ji_1\ldots i_{2n}}(\bm{w};\sigma^2)\\
% N_n^{-1}\mathbf{d}_{n}:\bm{h}_{|2n|+(2)}(\bm{w};\sigma^2)&=\sigma^{2(2n+2)}N_n^{-1}\mathbf{d}^{k\ell}_{n}(\delta_{i_1
%     i_2}\cdots\delta_{i_{2n-1}i_{2n}})\mg{2n+2}^{k\ell i_1\ldots i_{2n}}(\bm{w};\sigma^2)
% \end{align}
The coefficients $\bm{c}_{(i)}$ for a Hermite expansion of the
distribution function compatible with the Chapman-Enskog correction
equations are thus given, for $n\geq 1$, by
\begin{align}
%&\bm{c}_{(2n)}=(2n)!(\sigma^2)^{2n}\text{Sym}\left[\frac{1}{N_{n-1}}(\mathbf{d}_{n-1} \bm{\delta}_{2(n-1)} - \frac{1}{3}\text{tr}(\mathbf{d}_{n-1})\bm{\delta}_{2n})   
&\bm{c}_{(2n)}=(-1)^n\frac{(2n)!}{2^nn!}\sigma^{2n}\text{Sym}\left[a_n\bm{\delta}_{2n}
-2n\sigma^2(\mathbf{d}_{n-1} \bm{\delta}_{2n-2} - \frac{1}{3}\text{tr}[\mathbf{d}_{n-1}]\bm{\delta}_{2n})   \right]\\
%&\bm{c}_{(2n+1)}=(2n+1)!\sigma^{2(2n+1)}\text{Sym}\left[\frac{\bm{b}_n}{N_n}\bm{\delta}_{2n}\right]
&\bm{c}_{(2n+1)}=(-1)^n\frac{(2n+1)!}{2^nn!}\sigma^{2n+2}\text{Sym}\left[\bm{b}_n\bm{\delta}_{2n}\right].
\end{align}
where the coefficient $a_1 = 0$ is reminded to be zero and the
short-hand notation,
$\bm{\delta}_{2n}=\delta_{i_1i_2}\delta_{i_3i_4}\cdots\delta_{i_{2n-1}i_{2n}}$,
for the pair-wise contraction operator is used.  Essentially, the
result emerges from expressing spherically based tensor objects in
Cartesian coordinates \cite[Chapter 4]{Wein12}. These coefficients can
be used in (\ref{eq:drag_term}) and (\ref{eq:diffusion_term}) to
express the collisional moments of the Landau operator, where only the
linear terms would be retained to be consistent with the
Chapman-Enskog procedure.}
\section{Collisional ten-moment equations}
\label{sec:ten-moment}
\revision{The ten-moment equations represent the simplest possible
  projection of the kinetic equation beyond Maxwellian behaviour. In
  effect, they are derived by truncating the distribution functions
  after the second order Hermite polynomials. This way, a closed set
  of collisional fluid equations is obtained in which the stress
  tensor features as a dynamical moment on an equal footing to
  density, flow and temperature. In the limit of vanishing
  mean-free-path, only the concept of viscosity emerges from the
  ten-moment model. Conductivity comes in pair with heat flux, for
  example in Grad's 13-moment equations and higher-moment theories
  \cite[Chapter 4]{balescu-1988}. Nevertheless, the simple ten-moment
  model is useful to illustrate how to apply our formulae for the
  collisional moments of the Landau operator, as well as to
  demonstrate their conservation properties}. We thus proceed to
consider distribution functions of the form
  \begin{align}
  f_s(\bm{v}) &= n_s\mathcal{N}_{\sigma_s^2}(\bm{v}-\bm{V}_s)[1 + \tfrac{1}{2}\bm{c_{s(2)}} \mg{2}(\bm{v}-\bm{V}_s;\sigma_s^2)]\nonumber\\ 
&= n_s\left(\frac{m_s}{2\pi T_s}\right)^{3/2}e^{-\frac{m_s}{2T_s}(\bm{v}-\bm{V}_s)^2}\left[1 + \frac{m_s}{2T_s}(\bm{v}-\bm{V}_s)\cdot\left(\frac{\bm{P}_s}{p_s} - \bm{I}\right)\cdot(\bm{v}-\bm{V}_s)\right].
\end{align}
for which there are 10 variables for each species, namely the density
$n_s$ (1 scalar), mean velocity $\bm{V}_s$ (3-component vector) and
pressure tensor $\bm{P}_s$ (6-component symmetric matrix, including
$p_s=n_s T_s$ as its trace). The extended fluid equations are then
determined via the system
\begin{align}
\label{eq:density_eq}
\int_{\mathbb{R}^3}d\bm{v}\; m_s\frac{d f_s}{dt} &
\equiv C_{ss'(0)} %=0
\\
\label{eq:momentum_eq}
\int_{\mathbb{R}^3}d\bm{v}\; m_s\bm{v}\frac{d f_s}{dt} &
\equiv \sum_{s'}(\bm{C}_{ss'(1)}+\bm{V}_sC_{ss'(0)})
%= \sum_{s'}\bar{c}_{ss'}\mu_{ss'} \bm{R}_{ss'(1)} = \sum_{s'} \bm{F}_{ss'} 
\\
%\int_{\mathbb{R}^3}d\bm{v}\; v^2\frac{df_s}{dt} &
%\equiv \frac{\partial}{\partial t}\left(\int_{\mathbb{R}^3}d\bm{v}\; v^2f\right)+\nabla\cdot\left(\int_{\mathbb{R}^3}d\bm{v}\;\bm{v}v^2f\right)-\bm{E}\cdot\left(\int_{\mathbb{R}^3}d\bm{v}\; 2\bm{v}f\right) 
%\equiv \beta^{-2}_s\mathrm{Tr}\left[C^{ss'}_{(2)}\right]+2\beta^{-1}_s\bm{V}_s\cdot C^{ss'}_{(1)}+\left(V^2_s+3\beta_s^{-2}\right)C^{ss'}_{(0)}
%\\
\label{eq:pressure_eq}
\int_{\mathbb{R}^3}d\bm{v}\; m_s\bm{v} \bm{v}\frac{df_s}{dt} &
\equiv \sum_{s'}(\bm{C}_{ss'(2)}+\bm{V}_{s} \bm{C}_{ss'(1)}+\bm{C}_{ss'(1)} \bm{V}_{s} + \bm{V}_s\bm{V}_s C_{ss'(0)} + \sigma_s^2\bm{I}C_{ss'(0)})
\end{align}

\subsection{Moments of the Vlasov operator}
Considering that the Vlasov operator for plasmas, in the absence of gravitational forces, is given by
\begin{equation}
\frac{d}{dt}=\frac{\partial}{\partial t}+\bm{v}\cdot\nabla_{\bm{x}}+\frac{e_s}{m_s}(\bm{E}+\bm{v}\times\bm{B})\cdot\nabla_{\bm{v}},
\end{equation}
we may write the fluid equations
explicitly. Defining the mass density $\rho_s\equiv m_sn_s$ and the
momentum vector $\bm{K}_s\equiv \rho_s\bm{V}_s$, equation
(\ref{eq:density_eq}) corresponds to the familiar continuity equation
\begin{equation}
%\frac{\partial \rho_s}{\partial t}+\frac{\partial K_s^i}{\partial x^i}=0. \label{eq:mass_eq_exp}
\frac{\partial \rho_s}{\partial t}+\nabla\cdot\bm{K}_s=0, \label{eq:mass_eq_exp}
\end{equation}
since $C_{ss'(0)}=0$. Defining the stress tensor
$\bm{\Pi}_s\equiv \bm{P}_s+\rho_s\bm{V}_s\bm{V}_s$, equation
(\ref{eq:momentum_eq}) becomes the momentum equation
\begin{equation}
%\frac{\partial K_s^i}{\partial t}+\frac{\partial \Pi_s^{ij}}{\partial x^j}-\frac{e_s}{m_s}(\rho_sE^i+\varepsilon_{ijk}K_s^jB^k)= \sum_{s'}\bar{c}_{ss'}\mu_{ss'} R^i_{ss'(1)}, \label{eq:momentum_eq_exp}
\frac{\partial \bm{K}_s}{\partial t}+\nabla\cdot \bm{\Pi}_s-\frac{e_s}{m_s}(\rho_s\bm{E}+\bm{K}_s\times \bm{B})= \sum_{s'}\bar{c}_{ss'}\mu_{ss'} \bm{R}_{ss'(1)}, \label{eq:momentum_eq_exp}
\end{equation}
and equation (\ref{eq:pressure_eq}) provides the evolution equation for the stress tensor
\begin{multline}
\frac{\partial\Pi_s^{ij}}{\partial t}+\frac{\partial}{\partial x^k}\left(\frac{\Pi^{ij}K^k+\Pi^{jk}_sK_s^i+\Pi^{ki}K_s^j}{\rho_s}-2\frac{K_s^iK_s^jK_s^k}{\rho_s^2}\right)
%-\frac{e_s}{m_s}(E^i K_s^j+K^i_s E^j+B^m(\varepsilon_{i\ell m}\Pi_s^{j\ell}+\varepsilon_{j\ell m}\Pi_s^{i\ell}))
-\frac{e_s}{m_s}(E^i K_s^j+B^m\varepsilon_{i\ell m}\Pi_s^{j\ell} + \text{transpose})
\\=\sum_{s'}\bar{c}_{ss'}[m_s^{-1}D^{ij}_{ss'(2)}+\mu_{ss'}(R^{ij}_{ss'(2)}+V^i_{s} R^j_{ss'(1) }) + \text{transpose}]. \label{eq:stress_eq_exp}
\end{multline}
Rather than using the number density, mean velocity, and pressure tensor,
the mass density, the momentum vector, and the stress tensor were
respectively introduced for the sake of expressing the moment
equations in a divergence form; in numerical implementations of these
equations, the divergence form allows for the use of conservative
discretisation
methods. % The expressions for the stress tensor, momentum vector, and mass density have simple dependencies on the density, mean velocity, and pressure tensor
% \begin{equation}
% \bm{\Pi}_s=\bm{P}_s+m_sn_s\bm{V}_s\bm{V}_s, \quad \bm{K}_s=m_sn_s\bm{V}_s, \quad \rho_s=m_sn_s.
% \end{equation}

%
\subsection{Moments of the collision operator}
The collisional contributions to the ten-moment equations require the
determination of the coefficients $\bm{R}_{ss'(1)}$,
$\bm{R}_{ss'(2)}$, and $\bm{D}_{ss'(2)}$. By virtue
of~(\ref{eq:drag_term}) and~(\ref{eq:diffusion_term}) one readily
finds
\begin{align}
  \bm{R}_{ss'(1)} &=\frac{1}{2\Sigma^2_{ss'}}\nabla_{\bm{\Delta}_{ss'}}\mathcal{O}_{ss'}[\Phi](\Delta_{ss'}),\\
  \bm{R}_{ss'(2)}& = \frac{1}{\sqrt{2}\Sigma_{ss'}}\left[\frac{\sigma_s^2}{2\Sigma_{ss'}^2}\nabla_{\bm{\Delta}_{ss'}}\nabla_{\bm{\Delta}_{ss'}} \mathcal{O}_{ss'}[\Phi](\Delta_{ss'})
    + 2(\bm{\tilde{\pi}}_{s}\cdot\nabla_{\bm{\Delta}_{ss'}})\nabla_{\bm{\Delta}_{ss'}}\left(1 + \bm{\tilde{\pi}}_{s'}:\nabla_{\bm{\Delta}_{ss'}}\nabla_{\bm{\Delta}_{ss'}}\right)\Phi(\Delta_{ss'})\right],\\
  \bm{D}_{ss'(2)} &= \frac{1}{\sqrt{2}\Sigma_{ss'}}\nabla_{\bm{\Delta}_{ss'}}\nabla_{\bm{\Delta}_{ss'}}\mathcal{O}_{ss'}[\Psi](\Delta_{ss'}),
\end{align}
where the scalar differential operator ${\cal O}_{ss'}$ is given by
\begin{align}
\mathcal{O}_{ss'} &\equiv
1 + (\bm{\tilde{\pi}}_{s}+\bm{\tilde{\pi}}_{s'}):\nabla_{\bm{\Delta}_{ss'}}\nabla_{\bm{\Delta}_{ss'}} 
+ (\bm{\tilde{\pi}}_{s}:\nabla_{\bm{\Delta}_{ss'}}\nabla_{\bm{\Delta}_{ss'}})(\bm{\tilde{\pi}}_{s'}:\nabla_{\bm{\Delta}_{ss'}}\nabla_{\bm{\Delta}_{ss'}}).
\end{align}
and depends on the species $s$ and $s'$ via $\bm{\Delta}_{ss'}$ and
the normalised viscosity tensor,
\begin{equation}
\bm{\tilde{\pi}}_{s} = \frac{\bm{c}_{s(2)}}{4\Sigma_{ss'}^2} = \frac{1}{2}\frac{(\bm{P}_s-p_s\bm{I})/(m_sn_s)}{v^2_{th,s}+v^2_{th,s'}} % \frac{\bm{P}_s/p_s - \bm{I}}{4\left(1 + \frac{m_s T_{s'}}{m_{s'}T_s}\right)}.
\end{equation}
The operator ${\cal O}_{ss'}$ is observed to be symmetric with respect
to exchanging the species' indices. Thus, it is seen that
$\bm{R}_{ss'(1)}$ is antisymmetric while $\bm{D}_{ss'(2)}$ is
symmetric under this operation. This property remains true when
higher-moments are included.

\subsection{Conservation laws of the collisional moments}
The Landau collision operator conserves particle densities, total
kinetic momentum, and total kinetic energy. Since our procedure to
compute the collisional moments is exact, all fluid equations derived
by applying (\ref{eq:drag_term}) and (\ref{eq:diffusion_term})
automatically satisfy the same conservation properties, regardless of
the order of truncation. This statement is proven explicitly for the
ten-moment equations, although generalizations to higher moment fluid
theories are straightforward.

The conservation of particle densities is trivial because
$C_{ss'(0)} = 0$. The collisional momentum-transfer rate, given by
$\bm{F}_{ss'}=\bar{c}_{ss'}\mu_{ss'}\bm{R}_{ss'(1)}$, is
anti-symmetric with respect to changing the species indices, i.e.,
$\bm{F}_{ss'}=-\bm{F}_{ss'}$. This follows from the symmetry of the
operator ${\cal O}_{ss'}$ and the anti-symmetry of
$\nabla_{\bm{\Delta}_{ss'}}=-\nabla_{\bm{\Delta}_{s's}}$, thereby
establishing the conservation of total kinetic momentum.

%Momentum conservation is evident from the
%anti-symmetry of the friction force, obtained as the gradient with
%respect to $\bm{\Delta}_{ss'}=-\bm{\Delta}_{s's}$ of the symmetric
%scalar function
%$\bar{c}_{ss'}\mu_{ss'}\mathcal{O}_{ss'}[\Phi](\Delta_{ss'})/2\Sigma_{ss'}^2$.

The collisional energy-exchange rate to species $s$ from species $s'$
is defined as
\begin{align}
W_{ss'} = \frac{1}{2}\text{tr}\left[\int_{\mathbb{R}^3}d\bm{v}\, m_s \bm{v}\bm{v}\,C_{ss'}[f_s,f_{s'}]\right].
\end{align}
% Thus, we have
% \begin{multline}
% W_{ss'}+W_{s's}=\frac{1}{2}\bar{c}_{ss'}\text{tr}\left(2\,\text{Sym}\left[\mu_{ss'}(\bm{R}_{ss'(2)}+\bm{R}_{s's(2)})+\frac{1}{m_s}\bm{D}_{ss'(2)}+\frac{1}{m_{s'}}\bm{D}_{s's(2)}\right]\right)
% \\+\frac{1}{2}\bar{c}_{ss'}\mu_{ss'}\text{tr}\left(\bm{V}_s\bm{R}_{ss'(1)}+\bm{R}_{ss'(1)}\bm{V}_s+\bm{V}_{s'}\bm{R}_{s's(1)}+\bm{R}_{s's(1)}\bm{V}_{s'}\right).
% \end{multline}
The total energy-exchange rate can be expressed, thanks to the
symmetry of $\bm{D}_{ss'(2)}$ and the anti-symmetry of
$\bm{R}_{ss'(1)}$ with respect to species indices, as
\begin{align}
W_{ss'}+W_{s's}&=\bar{c}_{ss'}\mu_{ss'}\text{tr}\left[\bm{R}_{ss'(2)}+\bm{R}_{s's(2)}+\bm{D}_{ss'(2)} + \sqrt{2}\Sigma_{ss'}\bm{\Delta}_{ss'}\bm{R}_{ss'(1)}\right].
\end{align}
It is observed, after applying the identity $\nabla^{(n)} (\bm{x}\cdot
\nabla) = (\bm{x}\cdot\nabla)\nabla^{(n)} + n \nabla^{(n)}$, that
\begin{align}
\mathcal{O}_{ss'}[\bm{\Delta}_{ss'}\cdot\nabla_{\bm{\Delta}_{ss'}}]&=\bm{\Delta}_{ss'}\cdot\nabla_{\bm{\Delta}_{ss'}}{\cal O}_{ss'} +2(\bm{\tilde{\pi}}_{s} +\bm{\tilde{\pi}}_{s'}):\nabla_{\bm{\Delta}_{ss'}}\nabla_{\bm{\Delta}_{ss'}} \nonumber\\
&\quad+ 4(\bm{\tilde{\pi}}_{s}:\nabla_{\bm{\Delta}_{ss'}}\nabla_{\bm{\Delta}_{ss'}})(\bm{\tilde{\pi}}_{s'}:\nabla_{\bm{\Delta}_{ss'}}\nabla_{\bm{\Delta}_{ss'}}),
\end{align}
and thus the total energy-exchange rate vanishes identically by virtue
of
\begin{align}
W_{ss'}+W_{s's}
% =\frac{c_{ss'}\mu_{ss'}}{\sqrt{2}\Sigma_{ss'}}\biggr(\nabla_{\bm{\Delta}_{ss'}}\cdot\nabla_{\bm{\Delta}_{ss'}}{\cal O}_{ss'}[\Psi]+\bm{\Delta}_{ss'}\cdot\nabla_{\bm{\Delta}_{ss'}}{\cal O}_{ss'}[\Phi]+\frac{1}{2}\nabla_{\bm{\Delta}_{ss'}}\cdot\nabla_{\bm{\Delta}_{ss'}}{\cal O}_{ss'}[\Phi]\\+2(\bm{\tilde{\pi}}_{s} +\bm{\tilde{\pi}}_{s'}):\nabla_{\bm{\Delta}_{ss'}}\nabla_{\bm{\Delta}_{ss'}}\Phi + 4(\bm{\tilde{\pi}}_{s}:\nabla_{\bm{\Delta}_{ss'}}\nabla_{\bm{\Delta}_{ss'}})(\bm{\tilde{\pi}}_{s'}:\nabla_{\bm{\Delta}_{ss'}}\nabla_{\bm{\Delta}_{ss'}})\Phi\biggr)
% %\\
&=\frac{\bar{c}_{ss'}\mu_{ss'}}{\sqrt{2}\Sigma_{ss'}}{\cal O}_{ss'}\biggr[\nabla_{\bm{\Delta}_{ss'}}\cdot\nabla_{\bm{\Delta}_{ss'}}\Psi+\frac{1}{2}\nabla_{\bm{\Delta}_{ss'}}\cdot\nabla_{\bm{\Delta}_{ss'}}\Phi+\bm{\Delta}_{ss'}\cdot\nabla_{\bm{\Delta}_{ss'}}\Phi\biggr]
%&=\frac{\bar{c}_{ss'}\mu_{ss'}}{\sqrt{2}\Sigma_{ss'}}{\cal O}_{ss'}\biggr[\Phi-2\pi{\cal N}_{1/2}+\bm{\Delta}_{ss'}\cdot\nabla_{\bm{\Delta}_{ss'}}\Phi'\biggr]
%\nonumber\\
=0.
\end{align}
The last step depends on the properties of the functions $\Phi$
and $\Psi$, namely $\nabla\cdot\nabla\Psi = \Phi$ and
$\nabla\cdot\nabla\Phi = -4\pi\mathcal{N}_{1/2}(\bm{x}) =
-2\text{erf}'$ and
$\bm{x}\cdot\nabla\Phi = x \Phi' = \text{erf}' - \Phi$.
\section{Conclusion}
\label{sec:conclusion}
The formalism originally introduced by Grad to derive collisional
fluid theories has been applied to Coulomb interactions in warm
plasmas, with the sole (implicit) assumption that the Landau collision
operator is valid. As our main result, analytic expressions for the
collisional moments were obtained, that are represented fully in terms
of derivatives of the two Rosenbluth-MacDonald-Judd-Trubnikov
potentials for a Normal distribution with respect to the relative mean
flow normalised to the root-mean-square of the thermal
velocities. \revision{The formulae are manifestly Galilean invariant
  and highlight the nonlinear dependency of the collisional moments on
  local equilibrium Maxwellian variables, as well as the bilinearity
  with respect to higher-order moments of the distribution
  functions. Thanks to the correspondence between Laguerre and
  irreducible Hermite polynomials, a linearised version of our
  collisional moments can be used within the Chapman-Enskog approach
  to derive transport equations for plasmas and extend their validity
  to broader classes of flows, temperature differences and arbitrary
  species mass ratio}. Within Grad's approach, one can truncate the
expansion of the distribution function at a given polynomial degree,
and arrive at a fully closed, self-consistent system of extended fluid
equations. As an example, the ten-moment equations were presented, and
the collisional momentum- and energy-transfer rate were demonstrated
to preserve exact conservation properties, which is an often
overlooked requirement.

Employing our formulae to include collisional effects in fluid codes
will not only improve the physical accuracy of the modelling but will
also help alleviate numerical difficulties such as the build-up of
sharp gradients and formation of fine structures in a consistent and
controllable way. Another important consequence of our methodology is
that one can provide an exact dependency of the effective electrical
resistivity on all available fluid variables, not only on density and
temperature. The physical consequences of such resistivity tensor, and
its relevance in space and astrophysical plasmas (and fast magnetic
reconnection) is explored in the companion paper \cite{LHP16}.

Expressions for higher-order moments based on our formalism, although
more complex, are naturally programmable. The formulation could
possibly be widened to encompass also anisotropic distribution
functions and other non-Maxwellian features by generalising to
multivariate Gaussians and Hermite polynomials. Given the similarities
between the Coulomb and gravitational forces \cite{BinneyTremaine87},
we suggest that our methodology could also be applied to the
latter. Lastly, given the importance of the Hermite polynomials in the
quantum harmonic oscillator \citep{Wein12}, it is plausible that some
of the mathematical identities derived here could be utilised to study
this system.

\acknowledgments The authors would like to acknowledge useful
discussions with A.Bhattacharjee, J.A.Krommes and G.W.Hammett. The
insightful comments from the anonymous referee were greatly
appreciated. The authors were supported by the Department of Energy
Contract No. DE-AC02-09CH11466 during the course of this work.

\appendix
\section{Hermite polynomials: definitions and properties}
\label{sec:hermite_properties}
Many properties of the \emph{physicists'} Hermite polynomials,
$\mh{k}(\bm{x}) = \mg{k}(\bm{x};\frac{1}{2})$ (in our notation), were
originally derived by Grad \cite{Grad:1949_polynomials} and
successfully used in his work on the collisional moments of hard
spheres \cite{Grad:1949_rarefied_gas_theory}. The Landau collision
operator in plasmas being a much more complicated convolution, our
previous work \cite{hirvijoki-2016} relied on both the
\emph{physicists'} and \emph{probabilists'} Hermite polynomials,
$\mhe{k}(\bm{x})=\mh{k}(\bm{x};1)=\mg{k}(\bm{x};1)$, in order to
transfer the Gaussian convolution onto the Rosenbluth potentials. The
collisional moments were then generated by the gradients of three
scalar valued integrals, which proved to be quite tedious to
evaluate. The general approach of this paper is based on a wider class
of Hermite polynomials described by Holmquist \cite{Holmquist:1996}
and fully exploits their properties under convolution. In this
section, some fundamental identities from \cite{Holmquist:1996} are
reviewed and extended.
\paragraph{Contravariant Hermite polynomials:}
% According to Holmquist \citep[eq.2.1]{Holmquist:1996}, the
% contravariant Hermite polynomials are generated from the Normal
% distribution as
% \begin{equation}
%   \label{eq:hermite_def}
% \mhh{k}(\bm{x}-\bm{\mu};\sigma^2) = \frac{1}{\mathcal{N}_{\sigma^2}(\bm{x}-\bm{\mu})}(-\sigma^2\nabla_{\bm{x}})^k   \mathcal{N}_{\sigma^2}(\bm{x}-\bm{\mu}).
% \end{equation}
Owing to their definition in equation (\ref{eq:hermite_def}) or
\citep[eq.2.1]{Holmquist:1996}, the contravariant Hermite polynomials
can be obtained equivalently from
\begin{equation}
\label{eq:inv_weierstrass}
  \mhh{k}(\bm{x};\sigma^2)= e^{-\frac{\sigma^2}{2}\nabla_{\bm{x}}^2}\bm{x}^{(k)} = \left(\bm{x} - \sigma^2 \nabla_{\bm{x}}\right)^{(k)}1 =  \left(\bm{x} - \sigma^2 \nabla_{\bm{x}}\right) \mhh{k-1}(\bm{x};\sigma^2).
\end{equation}
The first few Hermite polynomials are listed
\begin{equation}
\label{eq:few_hermites}
\begin{aligned}
  \mhh{0}(\bm{x};\sigma^2)&= 1, &
  \mhh{1}(\bm{x};\sigma^2)&= \bm{x}, \\
  \mhh{2}(\bm{x};\sigma^2)&= \bm{x}\bm{x} - \sigma^2 \bm{I}, &
  \mhh{3}^{ijk}(\bm{x};\sigma^2)&= x^ix^jx^k - \sigma^2 (x^i \delta^{jk} + \delta^{ij}x^k + \delta^{ik}x^j).
\end{aligned}
\end{equation}
The contravariant Hermite polynomials \emph{scale} into each other
upon multiplication by a scalar $c$ \citep[eq.3.7]{Holmquist:1996} as
\begin{align}
\label{eq:scale_hermite}
c^k \mhh{k}(\bm{x};\sigma^2 )&=  \mhh{k}(c\bm{x};c^2\sigma^2 )  & \text{ or }&&
  \mhh{k}(c\bm{x};\sigma^2 ) &= c^k \mhh{k}(\bm{x};\sigma^2/c^2 ) &
\end{align}
\paragraph{Covariant polynomials:} The covariant Hermite polynomials
of equation (\ref{eq:covariant_gen}) or \citep[eq.3.8]{Holmquist:1996},
\begin{equation}
\label{eq:covariant_hermites_def}
  \mg{k}(\bm{x};\sigma^2 ) = \mhh{k}\left(\frac{\bm{x}}{\sigma^2}; \frac{1}{\sigma^2}\right)
    = \sigma^{-2k} \mhh{k}(\bm{x};\sigma^2),
\end{equation}
are conveniently used to express gradients of the Normal distribution
with respect to its mean as
\begin{align}
\label{eq:grad_gaussian}
\nabla_{\bm{\mu}}^{(k)}\mathcal{N}_{\sigma^2}(\bm{x}-\bm{\mu}) =
 (-\nabla_{\bm{x}})^{(k)}\mathcal{N}_{\sigma^2}(\bm{x}-\bm{\mu}) =
  \mg{k}(\bm{x}-\bm{\mu};\sigma^2)\mathcal{N}_{\sigma^2}(\bm{x}-\bm{\mu}).
\end{align}
% which are observed to be generated by the Normal distribution function
% by
% \begin{equation}
%   \label{eq:covariant_gen}
%   \mg{k}(\bm{x}-\bm{\mu};\sigma^2) = \frac{1}{\mathcal{N}_{\sigma^2}(\bm{x}-\bm{\mu})}(-\nabla_{\bm{x}})^k   \mathcal{N}_{\sigma^2}(\bm{x}-\bm{\mu}).
% \end{equation}
The first few covariant Hermite polynomials are listed
\begin{equation}
\label{eq:few_cov}
\begin{aligned}
  \mg{0}(\bm{x};\sigma^2)&= 1, &
  \mg{1}(\bm{x};\sigma^2)&= \frac{\bm{x}}{\sigma^2}, \\
  \mg{2}(\bm{x};\sigma^2)&= \frac{\bm{x}}{\sigma^2}\frac{\bm{x}}{\sigma^2} - \frac{\bm{I}}{\sigma^2}, &
  \mg{3}^{ijk}(\bm{x};\sigma^2)&= \frac{x^i}{\sigma^2}\frac{x^j}{\sigma^2}\frac{x^k}{\sigma^2} - \frac{1}{\sigma^4}(x^i \delta^{jk} + \delta^{ij}x^k + \delta^{ik}x^j)
\end{aligned}
\end{equation}
The scaling properties of covariant Hermite polynomials are opposite
to equation (\ref{eq:scale_hermite})
\begin{align}
\label{eq:scaling_hermite_covariant}
  \mg{k}(c\bm{x};c^2 \sigma^2) &= \mhh{k}\left(\frac{\bm{x}}{c \sigma^2};\frac{1}{c^2\sigma^2}\right) 
%= c^{-k} \mhh{k}\left(\frac{\bm{x}}{\sigma^2}; \frac{1}{\sigma^2}\right) 
= c^{-k}\mg{k}(\bm{x};\sigma^2) & \text{ or }&&
 c^{k}  \mg{k}(c\bm{x};c^2 \sigma^2) &= \mg{k}(\bm{x};\sigma^2).
\end{align}
\paragraph{Gauss-Weierstrass transform:}
Probably the most useful property of Hermite polynomials is that their
Gaussian mean (or Gaussian convolution) is a Hermite polynomial with a
different variance \citep[eq.9.5]{Holmquist:1996}
\begin{align}
\intR{x} \mhh{k}(\bm{x}-\bm{\nu};\sigma^2)\mathcal{N}_{\rho^2}(\bm{x}-\bm{\mu})  
= \mhh{k}(\bm{\mu}-\bm{\nu};\sigma^2-\rho^2)
\label{eq:cross_hermite}
\end{align}
provided that $\sigma^2\geq\rho^2$. For the covariant Hermite
polynomials, it reads
\begin{align}
\intR{x} \mg{k}(\bm{x}-\bm{\nu};\sigma^2)\mathcal{N}_{\rho^2}(\bm{x}-\bm{\mu})  
= \left(1-\frac{\rho^2}{\sigma^2}\right)^k\mg{k}(\bm{\mu}-\bm{\nu};\sigma^2-\rho^2).
\label{eq:cross_hermite_covariant}
\end{align}
The special case where $\sigma^2=\rho^2$, $\bm{x} \rightarrow \bm{y}$,
$\bm{\mu}=\bm{x}$ and $\bm{\nu}=\bm{0}$ in (\ref{eq:cross_hermite})
leads to the following expression for the Gauss-Weierstrass transform
of a Hermite polynomial
\begin{align}
\label{eq:weierstrass_hermite}
  \mathcal{G}_{\sigma^2}[\mhh{k}(\cdot;\sigma^2)](\bm{x}) 
  = \intR{y} \mhh{k}(\bm{y};\sigma^2)\mathcal{N}_{\sigma^2}(\bm{x}-\bm{y}) 
  = \mhh{k}(\bm{x};0) = \bm{x}^{(k)}
\end{align}
which is basically the inverse of (\ref{eq:inv_weierstrass}), proving
that the inverse Gauss-Weierstrass transform is
$\mathcal{G}^{-1}_{\sigma^2} \equiv e^{-\frac{1}{2}\sigma^2
  \nabla^2}$, as far as these expansions are concerned.
\paragraph{Convolution identities:}
It is often useful to split the product of two Gaussians in order to
isolate one of the variables ($\bm{y}$ in the case that follows). For
this, the quadratic exponent is arranged as
\begin{equation}
  \frac{1}{\sigma^2}(\bm{x}-\bm{y})^2 + \frac{1}{\tau^2} (\bm{y}-\bm{z})^2 
%= (A+B) \bm{y}^2 - 2 A \bm{x}\bm{y} + \frac{A^2}{A+B} \bm{x}^2 - \frac{A^2}{A+B}\bm{x}^2 + A \bm{x}^2 
= \frac{1}{\Gamma^2}\left(\bm{y} - \tfrac{\bm{x}+\bm{z}\sigma^2/\tau^2}{1+\sigma^2/\tau^2}\right)^2 + \frac{1}{\Sigma^2}(\bm{x}-\bm{z})^2
\end{equation}
where
\begin{align}
  \Sigma^2 &= \sigma^2 + \tau^2, &
 \Gamma^2 &=\frac{1}{\frac{1}{\sigma^2}+\frac{1}{\tau^2}}  = \frac{\sigma^2\tau^2}{\sigma^2+\tau^2}, &
\Gamma\Sigma &= \sigma\tau,
\end{align}
so that the product of two Gaussians becomes
\begin{align}
\label{eq:gauss_cut}
  \mathcal{N}_{\sigma^2}(\bm{x}-\bm{y})\mathcal{N}_{\tau^2}(\bm{y}-\bm{z})
%& = \frac{e^{-\frac{1}{2\sigma^2}(\bm{x}-\bm{y})^2-\frac{1}{2\tau^2}(\bm{y}-\bm{z})^2}}{(2\pi)^D\sigma^D\tau^D}
% =\frac{e^{-\frac{1}{2 \Gamma^2}(\bm{y}-\frac{\bm{x}+\bm{z}\sigma^2/\tau^2}{1+\sigma^2/\tau^2})^2}e^{-\frac{1}{2\Sigma^2}(\bm{x}-\bm{z})^2}}{(2\pi)^{D/2}\Gamma^D (2\pi)^{D/2}\Sigma^D}
 = \mathcal{N}_{\Gamma^2}\left(\bm{y}-\frac{\bm{x}+\bm{z}\sigma^2/\tau^2}{1+\sigma^2/\tau^2}\right)\mathcal{N}_{\Sigma^2}(\bm{x}-\bm{z}).
\end{align}
With this decomposition, one immediately shows that the convolution of
two Gaussians is a Gaussian with the sum of the variances
\begin{equation}
  \label{eq:convolution_gaussians}
  (\mathcal{N}_{\sigma^2} * \mathcal{N}_{\tau^2})(\bm{x}) = \intR{y}\mathcal{N}_{\sigma^2}(\bm{x}-\bm{y})\mathcal{N}_{\tau^2}(\bm{y})  = \mathcal{N}_{\Sigma^2}(\bm{x})
\end{equation}
More generally, the Gaussian convolution of a Normal distribution
times a Hermite polynomial is found to be
\begin{align}
\mathcal{G}_{\sigma^2}[\mathcal{N}_{\tau^2}(\cdot-\bm{z})& \mg{i}(\cdot-\bm{w};\rho^2) ](\bm{x}) =
 \intR{y} \mathcal{N}_{\sigma^2}(\bm{x}-\bm{y})\mathcal{N}_{\tau^2}(\bm{y}-\bm{z}) \mg{i}(\bm{y}-\bm{w};\rho^2) \nonumber\\
% &\overset{(\ref{eq:gauss_cut})}{=}  \mathcal{N}_{\Sigma^2}(\bm{x}-\bm{z}) \intR{y} \mathcal{N}_{\Gamma^2}(\bm{y}-\tfrac{\bm{x}+\bm{z}\sigma^2/\tau^2}{1+\sigma^2/\tau^2})\mg{i}(\bm{y}-\bm{w};\rho^2) \nonumber\\
% &\overset{(\ref{eq:cross_hermite_covariant})}{=} \mathcal{N}_{\Sigma^2}(\bm{x}-\bm{z})(1-\tfrac{\Gamma^2}{\rho^2})^i
% \mg{i}(\tfrac{\bm{x}+\bm{z}\sigma^2/\tau^2}{1+\sigma^2/\tau^2}-\bm{w} ; \rho^2-\Gamma^2)\nonumber\\
&\overset{(\ref{eq:gauss_cut}),(\ref{eq:cross_hermite_covariant}),(\ref{eq:scaling_hermite_covariant})}{=} \mathcal{N}_{\Sigma^2}(\bm{x}-\bm{z})
\mg{i}\left(\frac{\bm{x}-\bm{w}+(\bm{z}-\bm{w})\sigma^2/\tau^2}{1+\sigma^2/\tau^2 -\sigma^2/\rho^2};\frac{\Sigma^2}{\sigma^2/\rho^2+\tau^2/\rho^2-\sigma^2\tau^2/\rho^4}\right).\label{eq:weierstrass-covariant}
\end{align}
The latter formula is particularly elegant for the case where $\rho^2=\tau^2$,
\begin{align}
\label{eq:straight_beauty}
\mathcal{G}_{\sigma^2}[\mathcal{N}_{\tau^2}(\cdot-\bm{z}) \mg{i}(\cdot-\bm{w};\tau^2) ](\bm{x}) 
% \intR{y} \mathcal{N}_{\sigma^2}(\bm{x}-\bm{y})\mathcal{N}_{\tau^2}(\bm{y}-\bm{z}) \mg{i}(\bm{y}-\bm{w};\rho^2)\nonumber \\  
=\mathcal{N}_{\Sigma^2}(\bm{x}-\bm{z})
\mg{i}\left(\bm{x}-\bm{w}+(\bm{z}-\bm{w})\frac{\sigma^2}{\tau^2};\Sigma^2\right).
\end{align}
This identity was used in our previous work \cite{hirvijoki-2016} to
present the distribution function as a Gaussian convolution.
% and for $\rho^2=\sigma^2$,
% \begin{align}
% \mathcal{G}_{\sigma^2}[\mathcal{N}_{\tau^2}(\bm{y}-\bm{z}) \mg{i}(\bm{y}-\bm{w};\sigma^2) ](\bm{x}) 
% % \intR{y} \mathcal{N}_{\sigma^2}(\bm{x}-\bm{y})\mathcal{N}_{\tau^2}(\bm{y}-\bm{z}) \mg{i}(\bm{y}-\bm{w};\rho^2)\nonumber \\  
% =\mathcal{N}_{\Sigma^2}(\bm{x}-\bm{z})
% \mg{i}\left(\bm{z}-\bm{w}+(\bm{x}-\bm{w})\frac{\tau^2}{\sigma^2};\Sigma^2\right).
% \end{align}
%
% \paragraph{Gradient:}
% The directional derivative of a Hermite polynomial with respect to its
% argument is \citep[eq.6.2]{Holmquist:1996}
% \begin{equation}
% \label{eq:hermite_gradient}
%   \bm{J}\cdot\nabla_{\bm{x}} \mhh{k}(\bm{x};\sigma^2) 
% = J^i\partial_i  \mhh{k}^{a_1\cdots a_k}(\bm{x};\sigma^2) 
%   = \frac{1}{(k-1)!} J^{[a_1}\mhh{k-1}^{a_2\cdots a_{k}]}(\bm{x};\sigma^2) = k \text{ Sym}[ \bm{J} \mhh{k-1}(\bm{x};\sigma^2)]
% \end{equation}
% where $\text{Sym}A^{a_1\cdots a_k} = A^{[a_1 \cdots a_k]}/k!$ is the
% symmetrisation of a tensor. This result is consistent with Grad
% \citep[eq.17]{Grad:1949_polynomials}.
%
% =======
\paragraph{Linearisation:}
The product of two Hermite polynomials is conveniently
\emph{linearised} as
\begin{equation}\label{eq:linearization}
\mhh{i}(\bm{x};\sigma^2)\mhh{j}(\bm{x};\sigma^2)=\sum_{l=0}^{i+j}\bm{a}^{(l)}_{(i)(j)}\mg{l}(\bm{x};\sigma^2),
\end{equation}
where the linearisation coefficient is found to be (see appendix
\ref{sec:linearisation} for a detailed proof)
\begin{align}
\label{eq:neutral_coeff}
\bm{a}^{(l)}_{(i)(j)} &= \sigma^{i+j+l}\bm{\bar{a}}^{(l)}_{(i)(j)} &
\bm{\bar{a}}^{(l)}_{(i)(j)} &= \frac{1}{l!}\nabla_{\bm{x}}^{(i)}\nabla_{\bm{y}}^{(j)}\nabla_{\bm{z}}^{(l)} \left[e^{\bm{x}\cdot\bm{y} + \bm{y}\cdot\bm{z} + \bm{x}\cdot\bm{z}}\right]_{\bm{x}=\bm{0},\bm{y}=\bm{0},\bm{z}=\bm{0}}.
\end{align}
In particular, we evidently have
\begin{equation}
\label{eq:lincoef0}
  \bm{\bar{a}}^{(l)}_{(i)(0)} = \frac{1}{l!}\delta_{(i)}^{[(l)]}.
\end{equation}
We also note that for $i>0$
\begin{align}
  \bm{J}\cdot\mhh{1}(\bm{x};1)\mhh{i}(\bm{x};1) \overset{(\ref{eq:inv_weierstrass})}{=} \bm{J}\cdot\mhh{i+1}(\bm{x};1) + \bm{J}\cdot\nabla \mhh{i}(\bm{x};1)
\overset{(\ref{eq:hermite_gradient})}{=}\bm{J}\cdot \mhh{i+1}(\bm{x};1) + i \text{Sym}[\bm{J}\mhh{i-1}(\bm{x};1)]
\end{align}
i.e. it amounts to
\begin{align}
\label{eq:lincoef1}
  \bm{\bar{a}}^{(l)}_{(i)(1)} = \frac{1}{l!}\delta_{(i+1)}^{[(l)]} + \frac{1}{l!}\delta_{(i)}^{[(l+1)]}
\end{align}
which can also be obtained by using the second relation in
(\ref{eq:neutral_coeff}). The above identity comes in handy for
expressing (with $i>0$)
\begin{align}
\label{eq:reduclin}
  \sum_{l=0}^{i+1} \bm{c}_{(i)} \bm{\bar{a}}^{(l)}_{(i)(1)} \nabla^{(l)}
% &=\frac{1}{(i+1)!} c_{(i)}^{i_1\dots i_i} \delta^{l_1}_{[i_1}\dots\delta^{l_i}_{i_i}\delta^{l_{i+1}}_{m]} \partial_{l_1}\dots\partial_{l_{i+1}}
% + \frac{1}{(i-1)!}c_{(i)}^{i_1\dots i_i}\delta^{[l_1}_{i_1}\dots\delta^{l_{i-1}}_{i_{i-1}}\delta^{m]}_{i_i}\partial_{l_1}\dots\partial_{l_{i-1}}\\
% &=\frac{1}{(i+1)!} c_{(i)}^{i_1\dots i_i}\partial_{[i_1}\dots\partial_{i_i}\partial_{m]}
% + \frac{i}{i!}c_{(i)}^{[i_1\dots i_{i-1}m]}\partial_{i_1}\dots\partial_{i_{i-1}}\\
&= \bm{c}_{(i)}\nabla^{(i)}\nabla + i \bm{c}_{(i)}\nabla^{(i-1)}.
\end{align}

\section{Linearisation coefficient for products of Hermite polynomials}
\label{sec:linearisation}
The linearisation coefficient for the tensor product of two Hermite
polynomials is computed explicitly
\begin{align}\label{eq:linearization_coeff}
\bm{a}^{(k)}_{(i)(j)}&=\frac{1}{k!}\intR{x}\mhh{i}(\bm{x};\sigma^2)\mhh{j}(\bm{x};\sigma^2)\mhh{k}(\bm{x};\sigma^2) \mathcal{N}_{\sigma^2}(\bm{x})\nonumber\\
&\overset{(\ref{eq:covariant_hermites_def})}{=}\frac{\sigma^{2(i+j+k)}}{k!}\intR{x}\mg{i}(\bm{x};\sigma^2)\mg{j}(\bm{x};\sigma^2)\mg{k}(\bm{x};\sigma^2) \mathcal{N}_{\sigma^2}(\bm{x})\nonumber\\
&\overset{(\ref{eq:covariant_gen})}{=}\frac{\sigma^{2(i+j+k)}}{k!}\intR{x}\mg{i}(\bm{x};\sigma^2)\mg{j}(\bm{x};\sigma^2)\nabla_{\bm{u}}^{(k)} \mathcal{N}_{\sigma^2}(\bm{x}-\bm{u})\Big|_{\bm{u}=\bm{0}}\nonumber\\
&=\frac{\sigma^{2(i+j+k)}}{k!}\intR{x}\mg{i}(\bm{x};\sigma^2)\mg{j}(\bm{x};\sigma^2)\mathcal{N}_{\sigma^2}(\bm{x})\nabla_{\bm{u}}^{(k)}[e^{-\frac{1}{2\sigma^2}(\bm{u}^2-2\bm{x}\cdot\bm{u})}] \Big|_{\bm{u}=\bm{0}}\nonumber\\
&=\frac{\sigma^{2(i+j+k)}}{k!}\nabla_{\bm{w}}^{(i)}\nabla_{\bm{v}}^{(j)}\nabla_{\bm{u}}^{(k)}\intR{x} \frac{e^{-\frac{1}{2\sigma^2}[\bm{u}^2+\bm{v}^2+\bm{w}^2-2\bm{x}\cdot(\bm{u}+\bm{v}+\bm{w}) + \bm{x}^2]}}{(2\pi)^{3/2}\sigma^3}\Big|_{\bm{u}=\bm{0},\bm{v}=\bm{0},\bm{w}=\bm{0}}\nonumber\\
&=\frac{\sigma^{2(i+j+k)}}{k!}\nabla_{\bm{w}}^{(i)}\nabla_{\bm{v}}^{(j)}\nabla_{\bm{u}}^{(k)} e^{\frac{1}{\sigma^2}(\bm{u}\cdot\bm{v} + \bm{v}\cdot\bm{w} + \bm{u}\cdot\bm{w})} \cancelto{1}{\intR{x} \mathcal{N}_{\sigma^2}[\bm{x}-(\bm{u}+\bm{v}+\bm{w})] }\Big|_{\bm{u}=\bm{0},\bm{v}=\bm{0},\bm{w}=\bm{0}}\nonumber\\
&=\frac{\sigma^{i+j+k}}{k!}\nabla_{\bm{x}}^{(i)}\nabla_{\bm{y}}^{(j)}\nabla_{\bm{z}}^{(k)} \left[e^{\bm{x}\cdot\bm{y} + \bm{y}\cdot\bm{z} + \bm{x}\cdot\bm{z}}\right]_{\bm{x}=\bm{0},\bm{y}=\bm{0},\bm{z}=\bm{0}}=\sigma^{i+j+k}\bm{\bar{a}}^{(k)}_{(i)(j)}.
\end{align}
\revision{
\section{Correspondence between irreducible Hermite and Laguerre polynomials}
\label{app:laguerre}
It is useful to remark that maximally contracted multi-index Hermite
polynomials are proportional to a family of Laguerre polynomials. This
connection was noticed by Balescu \cite[appendix G]{balescu-1988} and
is essentially related to changing from Cartesian to spherical
coordinates. A formal proof of Balescu's construction is provided
here.

\paragraph{Laguerre polynomials: }
The generalised Laguerre polynomials are obtained via the following Rodrigues
formula
\begin{align}
  \label{eq:laguerre}
  L^\alpha_n ( y ) = \frac{e^y}{y^\alpha n!} \frac{d^n}{dy^n}(e^{-y}y^{n+\alpha}) 
%= \frac{1}{y^\alpha n!} \sum_{i=0}^n (-1)^i \frac{d^{n-i}}{dy^{n-i}}y^{n+\alpha}
= \sum_{i=0}^n (-1)^i {n+\alpha \choose n-i} \frac{y^i}{i!}
\end{align}
where the generalised binomial expansion,
\begin{align}
  \label{eq:binom_gene}
  { n + \alpha \choose m } \equiv \frac{(n+\alpha)(n+\alpha-1)\cdots(n+\alpha-m+1)}{m!},
\end{align}
and the Leibniz rule, $d^n(fg) = \sum_{i=0}^n{n\choose i}(d^if)
(d^{n-i} g)$, are used to carry out the last step. It can be shown
that they satisfy the following property under differentiation
\begin{align}
  \label{eq:laguerre_deriv}
  \frac{d^k}{d y^k} L^\alpha_n(y) =
  \begin{cases}
    (-1)^k L^{\alpha+k}_{n-k}(y) & k\leq n\\
    0 & \text{otherwise}
  \end{cases}
\end{align}
as well as the so-called three-point rules
\begin{align}
  L^\alpha_n(y) & =  L^{\alpha+1}_n(y) - L^{\alpha+1}_{n-1}(y)  \label{eq:laguerre_threepoint} \\
  n L^\alpha_n(y) &=(n+\alpha)L^\alpha_{n-1}(y)-y L^{\alpha+1}_{n-1}(y)   \label{eq:laguerre_threepointx}\\
  n L^\alpha_n(y) &=(\alpha + 1 - y)L^{\alpha+1}_{n-1}(y) - xL^{\alpha+2}_{n-2}(y) \label{eq:laguerre_threepoint_two}
\end{align}

\paragraph{Irreducible Hermite polynomials: }
A family of maximally contracted even-order Hermite polynomials, or
so-called irreducible Hermite polynomials, is derived via the relation
(\ref{eq:inv_weierstrass}) by exponentiation of the Laplace operator
$\nabla^2=\nabla\cdot\nabla$,
\begin{align}
  \label{eq:ihermite_def}
  &h_{|2n|}(x;\sigma^2) = \delta_{i_1
    i_2}\cdots\delta_{i_{2n-1}i_{2n}} \mhh{2n}^{i_1\ldots i_{2n}}(\bm{x};\sigma^2) = e^{-\frac{\sigma^2}{2}\nabla^2}x^{2n}
  %\\
%  \label{eq:ihermiteg_def}
%  &g_{|2n|}(x;\sigma^2) = \delta_{i_1
%    i_2}\cdots\delta_{i_{2n-1}i_{2n}} \mg{2n}^{i_1\ldots i_{2n}}(\bm{x};\sigma^2) = \sigma^{-4n}h_{|2n|}(x;\sigma^2)
\end{align}
where $x= |\bm{x}|$. Expressing the Laplacian in terms of the
normalised radial variable $y = x^2/2\sigma^2$, the above
definition becomes
\begin{align}
  \label{eq:essence_laguerre}
e^{-\frac{\sigma^2}{2}\nabla^2}\frac{ x^{2n}}{2^n \sigma^{2n}}  = e^{-\Delta}y^n 
= \sum_{k=0}^\infty (-1)^k \frac{\Delta^k}{k!} y^n
= (-1)^n n!\sum_{i=0}^n (-1)^i { n + \frac{1}{2} \choose n-i} \frac{y^i}{i!}
\end{align}
where the Laplacian operator $\Delta =
\frac{1}{y^{1/2}}\frac{d}{dy}y^{3/2}\frac{d}{dy}$ is defined with
respect to the variable $y$. The last equality in equation
(\ref{eq:essence_laguerre}) follows from the fact that $\Delta$ acts
as a ladder operator and decreases the powers of $y$ by one, i.e.
$\Delta (y^n) = n\left(n+\frac{1}{2}\right) y^{n-1}$, such that only
terms up to $n$ matter in the sum and the order of indices can be
reversed to $k=n-i$.

From (\ref{eq:ihermite_def}), (\ref{eq:essence_laguerre}) and
(\ref{eq:laguerre}), the following correspondence is made between the
even-order \emph{irreducible} Hermite polynomials and the
$1/2$-Laguerre basis,
\begin{align}
  \label{eq:ihermite}
  h_{|2n|}(x;\sigma^2) = N_n\;L_n^{1/2}\left(\frac{x^2}{2\sigma^2}\right)
\end{align}
where the conversion coefficient, $ N_n =(-1)^n2^n \sigma^{2n} n!$, is
identified.

A family of maximally contracted odd-order Hermite vectors is derived
from $h_{|2n|}$ by applying the recursion formula
(\ref{eq:inv_weierstrass}),
\begin{align}
  \label{eq:ihermite_vector}
  \bm{h}_{|2n|+(1)}(\bm{x};\sigma^2) &= (\bm{x}-\sigma^2\nabla_{\bm{x}}) h_{|2n|}(x;\sigma^2)
  %= \bm{x}\left( 1 - \frac{\sigma^2}{x} \frac{d}{dx}\right) h_{|2n|}(x;\sigma^2)
  = \bm{x} N_n\left( 1 - \frac{d}{dy}\right) L_n^{1/2}(y)
  \overset{(\ref{eq:laguerre_deriv})}{=} \bm{x} N_n\left(L_n^{1/2}(y) + L_{n-1}^{3/2}(y)\right)
 \overset{(\ref{eq:laguerre_threepoint})}{=} \bm{x} N_n L_n^{3/2}(y).
% \\
%   \bm{h}^{ji_1\ldots i_{2n}}_{|2n|+(1)}(\bm{x};\sigma^2) &= \sigma^{2(2n+1)}(\delta_{i_1
%     i_2}\cdots\delta_{i_{2n-1}i_{2n}})\mg{2n+1}^{ji_1\ldots i_{2n}}(\bm{x};\sigma^2)
\end{align}
A family of second-rank even-order Hermite tensors are generated by
repeating the recursion on $\bm{h}_{|2n|+(1)}$,
\begin{align}
  \label{eq:ihermite_tensor}
  \bm{h}_{|2n|+(2)}(\bm{x};\sigma^2)& = (\bm{x}-\sigma^2\nabla_{\bm{x}}) \bm{h}_{|2n|+(1)}(\bm{x};\sigma^2)
%  = N_n\left[(\bm{x}\bm{x} - \sigma^2\bm{I} ) L_n^{3/2}(y) - \bm{x} \sigma^2 \nabla_{\bm{x}}L_n^{3/2}(y)\right]\\
%  = N_n\left[(\bm{x}\bm{x} - \sigma^2\bm{I} ) L_n^{3/2}(y) - \bm{x} \bm{x} \frac{d}{dy} L_n^{3/2}(y)\right]\nonumber\\
  \overset{(\ref{eq:laguerre_deriv})}{=} 2\sigma^2N_n\left[\left(\frac{\bm{x}\bm{x}}{2\sigma^2} - \frac{1}{2}\bm{I} \right) L_n^{3/2}(y) + \frac{\bm{x}\bm{x}}{2\sigma^2}L_{n-1}^{5/2}(y)\right].
% \\
%   \bm{h}^{k\ell i_1\ldots i_{2n}}_{|2n|+(2)}(\bm{x};\sigma^2) &= \sigma^{2(2n+2)}(\delta_{i_1
%     i_2}\cdots\delta_{i_{2n-1}i_{2n}})\mg{2n+2}^{k\ell i_1\ldots i_{2n}}(\bm{x};\sigma^2)
\end{align}
It is easily shown from (\ref{eq:laguerre_threepoint_two}) that the
contraction of the expression above correctly corresponds to
$h_{|2n+2|}(x;\sigma^2)$. Hence, noting that
$N_{n+1} = - 2 \sigma^2 (n+1) N_n$, the following traceless object is
formed by considering
\begin{align}
  \label{eq:ihermite_tensor-contract}
  \bm{h}_{|2n|+(2)}(\bm{x};\sigma^2) - h_{|2n+2|}(x;\sigma^2)\frac{\bm{I}}{3}
  &\overset{(\ref{eq:ihermite})}{=} 2 \sigma^2 N_n\left[\frac{\bm{x}\bm{x}}{2\sigma^2} L_n^{5/2}(y) + \frac{\bm{I}}{3}\left( (n+1)L_{n+1}^{1/2}(y)- \frac{3}{2}L_n^{3/2}(y)\right)\right]\nonumber\\
%  &\overset{(\ref{eq:laguerre_threepoint})}{=} 2 \sigma^2 N_n\left[\frac{\bm{x}\bm{x}}{2\sigma^2} L_n^{5/2}(y) + \frac{\bm{I}}{3}\left( (n+1)L_{n+1}^{3/2}(y)- (n+1+\frac{3}{2})L_n^{3/2}(y)\right)\right]\\
  &\overset{(\ref{eq:laguerre_threepoint},(\ref{eq:laguerre_threepointx})}{=} N_n\left[\bm{x}\bm{x} - x^2\frac{\bm{I}}{3}\right]L_n^{5/2}\left(\frac{x^2}{2\sigma^2}\right).
\end{align}
The hierarchy of higher-order tensorial Hermite polynomials extends
beyond this point by successively applying the recursion formula. At
each level, the traceless subparts can be identified with the
corresponding Laguerre polynomial thanks to the three-point rules. A
one-to-one correspondence between the Hermite and Laguerre basis is
thus established, as in Balescu \cite[appendix G]{balescu-1988}.
}

%%% Local Variables:
%%% mode: latex
%%% TeX-master: "letter"
%%% End:

%
%\input{hermites_properties}
%\input{inside}
%\input{direct}
%\input{ten_moments}

\bibliographystyle{apsrev4-1}
\bibliography{bibfile}

\end{document}